\documentclass[12pt,a4paper]{article}
\pdfoutput=1
\usepackage{amsmath,amsfonts,amssymb,epsfig}
\usepackage{amsthm}
\usepackage{latexsym}
\usepackage{slashed}
\usepackage[width=0.9\textwidth,margin={30pt,30pt},font=small,labelfont=bf]{caption}
\numberwithin{equation}{section}

\usepackage{hyperref}
\usepackage{subfig}

\usepackage{xcolor}
\hypersetup{
    colorlinks,
    linkcolor={red!0!black},
    citecolor={blue!50!black},
    urlcolor={blue!80!black}
}

\usepackage{graphicx}
\usepackage{color}
\usepackage{bm}
\usepackage{multirow}
\usepackage{array}
\usepackage{xspace}

\newcommand{\newc}{\newcommand*}

\long\def\begincomment#1\endcomment{%
        \begingroup\sf\baselineskip12pt#1\endgroup}

\newc{\etal}{\textrm{et al.}} 
\newc{\eg}{\textrm{e.g.}} 
\newc{\ie}{\textrm{i.e.}}
\newc{\etc}{\textrm{etc}}
\newc\vs{\textrm{vs.}}
\newc{\cl}{\rm {C.L.}}

\newc{\ev}{\ensuremath{\,\mathrm{eV}}}
\newc{\kev}{\ensuremath{\,\mathrm{keV}}}
\newc{\mev}{\ensuremath{\,\mathrm{MeV}}}
\newc{\gev}{\ensuremath{\,\mathrm{GeV}}}
\newc{\tev}{\ensuremath{\,\mathrm{TeV}}}
\newc{\MeV}{\mev} 
\newc{\TeV}{\tev}
\newc{\invpb}{\ensuremath{/\text{pb}}}
\newc{\invfb}{\ensuremath{\,\text{fb}^{-1}}}
\newc\nb{\ensuremath{\,\mathrm{nb}}} \newc\pb{\ensuremath{\,\mathrm{pb}}} \newc\fb{\ensuremath{\,\mathrm{fb}}}
\newc\pc{\ensuremath{\,\mathrm{pc}}}
\newc\kpc{\ensuremath{\,\mathrm{kpc}}}
\newc\mpc{\ensuremath{\,\mathrm{Mpc}}}
\newc\ps{\ensuremath{\,\mathrm{ps}}} 
\newc\cmeter{\ensuremath{\,\mathrm{cm}}} 
\newc\meter{\ensuremath{\,\mathrm{m}}} 
\newc\kmeter{\ensuremath{\,\mathrm{km}}}
\newc\second{\ensuremath{\,\mathrm{s}}}
\newc\msecond{\ensuremath{\,\mathrm{ms}}}
\newc\nsecond{\ensuremath{\,\mathrm{ns}}}
\newc\psecond{\ensuremath{\,\mathrm{ps}}}

\newc{\chisqmin}{\ensuremath{\chi^2_{\mathrm{min}}}}
\newc{\Delchisq}{\ensuremath{\Delta\chi^2}}
\newc{\chisq}{\ensuremath{\chi^2}}
\newc{\like}{\ensuremath{\mathcal{L}}}

\newc\lsim{\ensuremath{\mathrel{\rlap{\lower4pt\hbox{\hskip1pt$\sim$}}\raise1pt\hbox{$<$}}}}
\newc\gsim{\ensuremath{\mathrel{\rlap{\lower4pt\hbox{\hskip1pt$\sim$}}\raise1pt\hbox{$>$}}}}
\newc{\VEV}[1]{\ensuremath{\langle #1 \rangle}}
\newc{\dl}{\ensuremath{\stackrel{\leftarrow}{D}}}
\newc{\dr}{\ensuremath{\stackrel{\rightarrow}{D}}}

\newc{\scr}[1]{\ensuremath{\mathcal{#1}}}

\newc{\bcenter}{\begin{center}}    \newc{\ecenter}{\end{center}}
\newc{\bfl}{\begin{flushleft}}    \newc{\efl}{\end{flushleft}}
\newc{\bfr}{\begin{flushright}}    \newc{\efr}{\end{flushright}}

\newc{\bi}{\begin{itemize}}
\newc{\ei}{\end{itemize}}
\newc{\bed}{\begin{description}}
\newc{\eed}{\end{description}}
\newc{\ben}{\begin{enumerate}}
\newc{\een}{\end{enumerate}}

\newc{\be}{\begin{equation}}
\newc{\ee}{\end{equation}}
\newc{\bea}{\begin{eqnarray}}
\newc{\eea}{\end{eqnarray}}
\newc{\ra}{\rightarrow}

\newc{\alphas}{\ensuremath{\alpha_s}}
\newc{\alphatwo}{\ensuremath{\alpha_2}}
\newc{\alphaone}{\ensuremath{\alpha_1}}
\newc{\alphai}[1]{\ensuremath{\alpha_{#1}}}
\newc{\alphaem}{\ensuremath{\alpha_{\mathrm{em}}}}
\newc{\alphaeff}{\ensuremath{\alpha_{\mathrm{eff}}}}
\newc{\sineff}{\ensuremath{\sin \theta_{\mathrm{eff}}}}
\newc{\sinsqeff}{\ensuremath{\sin^2 \theta_{\mathrm{eff}}}}
\newc{\dalphahad}{\ensuremath{\Delta \alpha_{\mathrm{had}}}}
\newc{\yt}{\ensuremath{h_t}} \newc{\yb}{\ensuremath{h_b}} \newc{\ytau}{\ensuremath{h_{\tau}}}
\newc\mz{\ensuremath{M_Z}} 
\newc\mw{\ensuremath{m_W}}
\newc\mZ{\mz}        \newc\mW{\mw}
\newc\mhsm{\ensuremath{ m_{H_{\mathrm{SM}}}}}
\newc{\mtop}{\ensuremath{ m_t}}               \newc{\mtpole}{\ensuremath{ M_t}}
\newc{\mbottom}{\ensuremath{ m_b}} 
\newc{\mtau}{\ensuremath{ m_{\tau}}}
\newc{\mt}{\mtpole}
\newc{\mb}{\mbottom} 
\newc{\rgg}{\ensuremath{R_{h}(\gamma\gamma)}}
\newc{\rzz}{\ensuremath{R_{h}(ZZ)}}
\newc{\rtwogg}{\ensuremath{R_{h_2}(\gamma\gamma)}}
\newc{\rtwozz}{\ensuremath{R_{h_2}(ZZ)}}
\newc{\ronegg}{\ensuremath{R_{h_1}(\gamma\gamma)}}
\newc{\ronezz}{\ensuremath{R_{h_1}(ZZ)}}
\newc{\rsiggg}{\ensuremath{R_{h_\textrm{sig}}(\gamma\gamma)}}
\newc{\rsigzz}{\ensuremath{R_{h_\textrm{sig}}(ZZ)}}
\newc{\llbar}{\ensuremath{\ell\bar{\ell}}}
\newc{\tauptaum}{\ensuremath{ \tau^+\tau^-}}
\newc{\qqbar}{\ensuremath{ q\bar{q}}} \newc{\ppbar}{\ensuremath{ p\bar{p}}}
\newc{\bbbar}{\ensuremath{ b\bar{b}}} \newc{\ttbar}{\ensuremath{ t\bar{t}}}
\newc{\ffbar}{\ensuremath{ f\bar{f}}} \newc{\tautaubar}{\ensuremath{ \tau\bar{\tau}}}
\newc{\mchi}{\ensuremath{m_{\chi}}}
\newc{\squark}{\ensuremath{\tilde{q}}}
\newc{\slepton}{\ensuremath{\tilde{l}}}
\newc{\gluino}{\ensuremath{\tilde{g}}} 
\newc{\mgluino}{\ensuremath{{m_{\gluino}}}}
\newc{\tone}{\ensuremath{{\tilde{t}_1}}}

\newc{\sthw}{\ensuremath{ \sin\theta_W}}              \newc{\cthw}{\ensuremath{\cos\theta_W}}
\newc{\tanthw}{\ensuremath{ \tan\theta_W}}              \newc{\cotthw}{\ensuremath{\cot\theta_W}}
\newc{\ssqthw}{\ensuremath{\sin^2 \theta_W}}
\newc{\msbar}{\ensuremath{\overline{MS}}} \newc{\drbar}{\ensuremath{\overline{DR}}}
\newc{\mtmtsmmsbar}{\ensuremath{ m_t(m_t)^{\msbar}_{{\mathrm{SM}}}}}
\newc{\mtmtsmdrbar}{\ensuremath{ m_t(m_t)^{\drbar}_{{\mathrm{SM}}}}}
\newc{\mtmtmssmdrbar}{\ensuremath{ m_t(m_t)^{\drbar}_{{\mathrm{SUSY}}}}}
\newc{\mbmbmsbar}{\ensuremath{ m_b(m_b)^{\msbar} }}
\newc{\mbmbsmmsbar}{\ensuremath{ m_b(m_b)^{\msbar}_{{\mathrm{SM}}}}}
\newc{\mbmzsmmsbar}{\ensuremath{ m_b(\mz)^{\msbar}_{{\mathrm{SM}}}}}
\newc{\mbmzsmdrbar}{\ensuremath{ m_b(\mz)^{\drbar}_{{\mathrm{SM}}}}}
\newc{\mbmzmssmdrbar}{\ensuremath{ m_b(\mz)^{\drbar}_{{\mathrm{SUSY}}}}}
\newc{\mtaumzsmmsbar}{\ensuremath{ m_{\tau}(\mz)^{\msbar}_{{\mathrm{SM}}}}}
\newc{\mtaumzsmdrbar}{\ensuremath{ m_{\tau}(\mz)^{\drbar}_{{\mathrm{SM}}}}}
\newc{\mtaumzmssmdrbar}{\ensuremath{ m_{\tau}(\mz)^{\drbar}_{{\mathrm{SUSY}}}}}
\newc{\alphasmzms}{\ensuremath{\alpha_s(M_Z)^{\overline{MS}}}}
\newc{\alphaimzms}[1]{\ensuremath{\alpha_{#1}(M_Z)^{\overline{MS}}}}
\newc{\alphaemmz}{\ensuremath{\alpha_{\mathrm{em}}(M_Z)^{\overline{MS}}}}

\newc{\mzero}{\ensuremath{{m_0}}}
\newc{\mhalf}{\ensuremath{ m_{1/2}}}
\newc{\tanb}{\ensuremath{\tan\beta}}
\newc{\azero}{\ensuremath{ A_0}}
\newc{\bzero}{\ensuremath{ B_0}}
\newc{\signmu}{\ensuremath{\rm{sgn}\,\mu}}
\newc{\mueff}{\ensuremath{\mu_{\rm{eff}}}}
\newc{\lam}{\ensuremath{{\lambda}}}
\newc{\kap}{\ensuremath{{\kappa}}}
\newc{\alam}{\ensuremath{{A_{\lambda}}}}
\newc{\akap}{\ensuremath{{A_{\kappa}}}}
\newc{\hs}{\ensuremath{ H_s}}      
\newc{\mhs}{\ensuremath{ m_{H_s}}} 
\newc{\mgut}{\ensuremath{ M_{\rm GUT}}}
\newc{\mplanck}{\ensuremath{ M_{\rm P}}}      \newc{\mpl}{\ensuremath{ M_{\rm Pl}}}
\newc{\msusy}{\ensuremath{ M_{\rm SUSY}}}      \newc{\ms}{\ensuremath{ M_{\rm S}}}
 \newc{\mhl}{\ensuremath{m_\hl}} 
 \newc{\mhone}{\ensuremath{m_{h_1}}} 
 \newc{\mhtwo}{\ensuremath{m_{h_2}}} 
 \newc{\mglu}{\ensuremath{m_{\tilde g}}} 
 \newc{\mul}{\ensuremath{m_{\tilde{u}_L}}} 
 \newc{\mtone}{\ensuremath{m_{\tilde{t}_1}}} 
 \newc{\ma}{\ensuremath{m_A}} 
 \newc{\maone}{\ensuremath{m_{a_1}}} 
 \newc{\matwo}{\ensuremath{m_{a_2}}}
 \newc{\hone}{\ensuremath{h_1}}
 \newc{\htwo}{\ensuremath{h_2}}
 \newc{\aone}{\ensuremath{a_1}}
 \newc{\atwo}{\ensuremath{a_2}}
 \newc{\mhu}{\ensuremath{ m_{H_u}}}       
 \newc{\mhd}{\ensuremath{ m_{H_d}}}
 \newc{\mhusq}{\ensuremath{ m_{H_u}^2}}       
 \newc{\mhdsq}{\ensuremath{ m_{H_d}^2}}
 \newc{\mhuew}{\ensuremath{ m^{\ast}_{H_u}}}       
 \newc{\mhdew}{\ensuremath{ m^{\ast}_{H_d}}}
 \newc{\mhuewsq}{\ensuremath{ m^{\ast\, 2}_{H_u}}}       
 \newc{\mhdewsq}{\ensuremath{ m^{\ast\, 2}_{H_d}}}
 \newc{\hu}{\ensuremath{ H_u}}       
 \newc{\hd}{\ensuremath{ H_d}}
 \newc{\barmhu}{\ensuremath{ \bar{m}_{H_u}}}
 \newc{\barmhd}{\ensuremath{ \bar{m}_{H_d}}}

 \newc{\mqthree}{\ensuremath{m_{\widetilde{Q}_3}^2}}
 \newc{\muthree}{\ensuremath{m_{\tilde{u}_3}^2}}
 \newc{\mdthree}{\ensuremath{m_{\tilde{d}_3}^2}}
 \newc{\mlthree}{\ensuremath{m_{\widetilde{L}_3}^2}}
 \newc{\methree}{\ensuremath{m_{\tilde{e}_3}^2}}
 \newc{\mqtwo}{\ensuremath{m_{\widetilde{Q}_2}^2}}
 \newc{\mutwo}{\ensuremath{m_{\tilde{u}_2}^2}}
 \newc{\mdtwo}{\ensuremath{m_{\tilde{d}_2}^2}}
 \newc{\mltwo}{\ensuremath{m_{\widetilde{L}_2}^2}}
 \newc{\metwo}{\ensuremath{m_{\tilde{e}_2}^2}}
 \newc{\mqone}{\ensuremath{m_{\widetilde{Q}_1}^2}}
 \newc{\muone}{\ensuremath{m_{\tilde{u}_1}^2}}
 \newc{\mdone}{\ensuremath{m_{\tilde{d}_1}^2}}
 \newc{\mlone}{\ensuremath{m_{\widetilde{L}_1}^2}}
 \newc{\meone}{\ensuremath{m_{\tilde{e}_1}^2}}
 \newc{\msmul}{\ensuremath{m_{\tilde{\mu}_L}}}
 \newc{\msmur}{\ensuremath{m_{\tilde{\mu}_R}}}
 \newc{\msneumu}{\ensuremath{m_{\tilde{\nu}_{\mu}}}}
 \newc{\mone}{\ensuremath{M_1}}
 \newc{\monesq}{\ensuremath{M_1^2}}
 \newc{\mtwo}{\ensuremath{M_2}}
 \newc{\mtwosq}{\ensuremath{M_2^2}}
 \newc{\mthree}{\ensuremath{M_3}}
 \newc{\mthreesq}{\ensuremath{M_3^2}}
 \newc{\atau}{\ensuremath{{A_{\tau}}}}
 \newc{\at}{\ensuremath{{A_{t}}}}
 \newc{\ab}{\ensuremath{{A_{b}}}}
 \newc{\atausq}{\ensuremath{{A_{\tau}^2}}}
 \newc{\atsq}{\ensuremath{{A_{t}^2}}}
 \newc{\absq}{\ensuremath{{A_{b}^2}}}

 \newc{\dmzero}{\ensuremath{\Delta{_{m_0}}}}
 \newc{\dmhalf}{\ensuremath{\Delta{_{m_{1/2}}}}}
 \newc{\dmu}{\ensuremath{\Delta{_{\mu}}}}

 \newc{\pten}{\ensuremath{\psi_{10}}}
 \newc{\ffive}{\ensuremath{\phi_{5}}}
 \newc{\hfive}{\ensuremath{h_{5}}}
 \newc{\hbfive}{\ensuremath{h_{\bar{5}}}}
 \newc{\thet}{\ensuremath{\theta_{50}}}
 \newc{\thetb}{\ensuremath{\theta_{\,\overline{50}}}}
 \newc{\ptenhat}{\ensuremath{\hat{\psi}_{10}}}
 \newc{\ffivehat}{\ensuremath{\hat{\phi}_{5}}}
 \newc{\hfivehat}{\ensuremath{\hat{h}_{5}}}
 \newc{\hbfivehat}{\ensuremath{\hat{h}_{\bar{5}}}}
 \newc{\thethat}{\ensuremath{\hat{\theta}_{50}}}
 \newc{\thetbhat}{\ensuremath{\hat{\theta}_{\,\overline{50}}}}
 \newc{\si}{\ensuremath{\Sigma}}
 \newc{\mfive}{\ensuremath{m_5^2}}
 \newc{\mten}{\ensuremath{m_{10}^2}}
 \newc{\dfive}{\ensuremath{\Delta^2_5}}
 \newc{\dbfive}{\ensuremath{\Delta^2_{\bar{5}}}}
 \newc{\dfifty}{\ensuremath{\Delta^2_{50}}}
 \newc{\dfiftyb}{\ensuremath{\Delta^2_{\,\overline{50}}}}
 \newc{\msi}{\ensuremath{m_{\Sigma}^2}}
 \newc{\lamh}{\ensuremath{\lambda_{H}}}
 \newc{\lamhb}{\ensuremath{\lambda_{\bar{H}}}}
 \newc{\ah}{\ensuremath{A_{H}}}
 \newc{\ahb}{\ensuremath{A_{\bar{H}}}}
 \newc{\lams}{\ensuremath{\lambda_{S}}}
 \newc{\as}{\ensuremath{A_{S}}}
 \newc{\lamsig}{\ensuremath{\lambda_{\si}}}
 \newc{\asig}{\ensuremath{A_{\si}}}

 \newc{\msten}{\ensuremath{m_{16}^2}}
 \newc{\mhun}{\ensuremath{m_{126}^2}}
 \newc{\mhunb}{\ensuremath{m_{\bar{126}}^2}}
 \newc{\mthun}{\ensuremath{m_{210}^2}}
 \newc{\ahun}{\ensuremath{A_{\bar{126}}}}
 \newc{\yhun}{\ensuremath{Y_{\bar{126}}}}
 \newc{\aten}{\ensuremath{A_{10}}}
 \newc{\yten}{\ensuremath{Y_{10}}}
 \newc{\alone}{\ensuremath{A_{\lambda_1}}}
 \newc{\altwo}{\ensuremath{A_{\lambda_2}}}
 \newc{\althree}{\ensuremath{A_{\lambda_3}}}
 \newc{\althreeb}{\ensuremath{A_{\bar{\lambda_3}}}}
 \newc{\lone}{\ensuremath{\lambda_1}}
 \newc{\ltwo}{\ensuremath{\lambda_2}}
 \newc{\lthree}{\ensuremath{\lambda_3}}
 \newc{\lthreeb}{\ensuremath{\bar{\lambda_3}}}

\newc{\sigsip}{\ensuremath{\sigma^{\rm SI}_{p}}}	\newc{\sigsin}{\ensuremath{\sigma^{\rm SI}_{n}}}
\newc{\sigsdp}{\ensuremath{\sigma^{\rm SD}_{p}}}	\newc{\sigsdn}{\ensuremath{\sigma^{\rm SD}_{n}}}
\newc{\sigsi}{\ensuremath{\sigma^{\rm SI}}}	\newc{\sigsd}{\ensuremath{\sigma^{\rm SD}}}
\newc{\sigv}{\ensuremath{\sigma v}}

\newc{\avsigv}{\ensuremath{\langle \sigma v \rangle}}

\newc{\abund}{\ensuremath{ \Omega h^2}}
\newc{\omegadm}{\ensuremath{ \Omega_{{\rm DM}}}}     \newc{\abunddm}{\ensuremath{ \Omega_{{\rm DM}} h^2}} 
\newc{\omegam}{\ensuremath{ \Omega_{{\rm m}}}}       \newc{\abundm}{\ensuremath{ \Omega_{{\rm m}} h^2}}
\newc{\omegab}{\ensuremath{ \Omega_{{\rm b}}}}	\newc{\abundb}{\ensuremath{ \Omega_{{\rm b}} h^2}}
\newc{\omegatot}{\ensuremath{ \Omega_{{\rm TOT}}}}
\newc{\omegacdm}{\ensuremath{ \Omega_{{\rm CDM}}}}   \newc{\abundcdm}{\ensuremath{ \Omega_{{\rm CDM}} h^2}}
\newc{\omegalambda}{\ensuremath{ \Omega_{\Lambda}}} \newc{\abundlambda}{\ensuremath{ \Omega_{\Lambda} h^2}}
\newc{\omegarad}{\ensuremath{ \Omega_{{\rm rad}}}}  \newc{\abundrad}{\ensuremath{ \Omega_{{\rm rad}} h^2}}
\newc{\rhocrit}{\ensuremath{ \rho_{\rm crit}}}
\newc{\rhochi}{\ensuremath{ \rho_{\chi}}}
\newc{\abunchi}{\ensuremath{\Omega_\chi h^2}}
\newc{\abunS}{\ensuremath{\Omega_S h^2}}
\newc{\abundlsp}{\ensuremath{\Omega_{\rm LSP}h^2}}

\newc{\amu}{\ensuremath{ a_{\mu}}}        \newc{\amususy}{\ensuremath{ a_{\mu}^{\mathrm{SUSY}}}}
\newc{\amuexpt}{\ensuremath{ a_{\mu}^{\mathrm{expt}}}}        \newc{\amusm}{\ensuremath{ a_{\mu}^{\mathrm{SM}}}}
\newc\deltaamu{\ensuremath{\Delta a_{\mu}}} \newc{\deltaamususy}{\ensuremath{\delta a_{\mu}^{\mathrm{SUSY}}}}
\newc\gmtwo{\ensuremath{ (g-2)_{\mu}}} 
\newc{\deltagmtwomususy}{\ensuremath{\delta\left(g-2\right)_{\mu}^{\mathrm{SUSY}}}}
\newc{\deltagmtwomu}{\ensuremath{\delta\left(g-2\right)_{\mu}}}
\newc\BR{\ensuremath{\textrm{BR}}}
\newc\bsgamma{\ensuremath{ b\rightarrow s \gamma }}
\newc\bxsgamma{\ensuremath{\overline{B}\rightarrow X_{s}\gamma}}
\newc\brbsgamma{\ensuremath{\BR\left(\bsgamma\right)}}
\newc\brbxsgamma{\ensuremath{\BR\left(\bxsgamma\right)}}
\newc\bsmumu{\ensuremath{B_s\to\mu^+\mu^-}}
\newc\brbsmumu{\ensuremath{\BR\left(B_s\to\mu^+\mu^-\right)}}
\newc\bdmmumu{\ensuremath{\overline{B}_d\to\mu^+\mu^-}}
\newc\bbbarmix{\ensuremath{\overline{B}_s\mbox{-}B_s}}      
\newc\delmbs{\ensuremath{\Delta M_{B_s}}}
\newc{\butaunu}{\ensuremath{B_u \rightarrow \tau \nu}}
\newc{\brbutaunu}{\ensuremath{\BR\left(B_u \rightarrow \tau \nu\right)}}

\newc{\brmuegamma}{\ensuremath{\BR\left(\mu^{\pm}\rightarrow e^{\pm}\gamma\right)}}
\newc{\brtauegamma}{\ensuremath{\BR\left(\tau^{\pm}\rightarrow e^{\pm}\gamma\right)}}
\newc{\brtaumugamma}{\ensuremath{\BR\left(\tau^{\pm}\rightarrow \mu^{\pm}\gamma\right)}}
\newc{\brmuthreee}{\ensuremath{\BR\left(\mu^{\pm}\rightarrow e^{\pm}e^+e^-\right)}}
\newc{\brtauthreee}{\ensuremath{\BR\left(\tau^{\pm}\rightarrow e^{\pm}e^+e^-\right)}}
\newc{\brtauthreemu}{\ensuremath{\BR\left(\tau^{\pm}\rightarrow \mu^{\pm}\mu^+\mu^-\right)}}

\newcommand*{\SARAH}{SARAH}

\newcommand*{\micromegas}{MicrOMEGAs}

\let\oldcite\cite
\renewcommand*{\cite}{~\oldcite}

\newcommand*{\hl}{\ensuremath{h}}

\newc{\glzmu}{\ensuremath{{g^{Z\mu \mu}_{L}}}}
\newc{\grzmu}{\ensuremath{{g^{Z\mu \mu}_{R}}}}
\newc{\glwmu}{\ensuremath{{g^{W\mu \nu_\mu}_{L}}}}
\newc{\grwmu}{\ensuremath{{g^{W\mu \nu_\mu}_{R}}}}
\newc{\glzmuSM}{\ensuremath{{g^{Z\mu \mu}_{L,\textrm{SM}}}}}
\newc{\grzmuSM}{\ensuremath{{g^{Z\mu \mu}_{R,\textrm{SM}}}}}
\newc{\glwmuSM}{\ensuremath{{g^{W\mu \nu_\mu}_{L,\textrm{SM}}}}}
\newc{\grwmuSM}{\ensuremath{{g^{W\mu \nu_\mu}_{R,\textrm{SM}}}}}

\def\pdf{\textit{pDF}}
\def\cs{\textit{CS}}
\def\piz{\ensuremath{{\pi^0}}}

\def\a{\textbf{(a)}}
\def\b{\textbf{(b)}}

\def\epem{ e^+ e^-}

\def\udark{U(1)_D}
\def\gzp{g_{V}}
\def\aem{\alpha_{\rm em}}
\def\azp{\alpha_{D}}

\def\mzp{M_{V}}
\def\ms{M_{S}}
\def\mdm{M_{\chi}}
\def\eps{\varepsilon}

\newcommand{\ap}{V}

\def \neff{N_\mathrm{eff}}
\def \csv{\langle\sigma v\rangle}

\def\nn{\nonumber}
\begin{document}

\title{\Large  {\bf Light dark Higgs boson in minimal \\[0.3em] sub-GeV dark matter  scenarios}}

\author{\\ \let\thefootnote\relax Luc Darm\'e,$^{1,a}$\footnote{$^a$  \url{luc.darme@ncbj.gov.pl}} \let\thefootnote\relax Soumya Rao,$^{1,b}$\footnote{$^b$ \url{soumya.rao@ncbj.gov.pl}} and Leszek Roszkowski$^{1,2,c}$\footnote{ $^c$ \url{leszek.roszkowski@ncbj.gov.pl}}\\[5ex]
\small {\em $^1$ National Centre for Nuclear Research,}\\
\small {\em Ho{\. z}a 69, 00-681 Warsaw, Poland }\\
\small {\em$^2$ Consortium for Fundamental Physics, Department of Physics and Astronomy, } \\
\small {\em University of Sheffield, Sheffield S3 7RH, United Kingdom} 
}

\date{}
\maketitle

\abstract{Minimal scenarios with light (sub-GeV) dark matter whose relic density is
obtained from thermal freeze-out must include new light mediators. In particular, a very
well-motivated case is that of a new ``dark'' massive vector gauge boson mediator. The
mass term for such mediator is most naturally obtained by a ``dark Higgs mechanism'' which
leads to the presence of an often long-lived dark Higgs boson whose mass scale is the same
as that of the mediator. We study the phenomenology and experimental constraints on two
minimal, self-consistent dark sectors that include such a light dark Higgs boson. In one
the dark matter is a pseudo-Dirac fermion, in the other a complex scalar. We find that the
constraints from BBN and CMB are considerably relaxed in the framework of such minimal
dark sectors. We present detection prospects for the dark Higgs boson in existing and
projected proton beam-dump experiments. We show that future searches at experiments like
Xenon1T or LDMX can probe all the relevant parameter space, complementing the various
upcoming indirect constraints from astrophysical observations.}

\newpage 

\tableofcontents

\section{Introduction}

Among the many puzzles facing the Standard Model (SM) of particle physics, the issue of dark
matter (DM) is certainly one of the most pressing. While the prime candidate of the last decades has
been the Weakly Interactive Massive Particle (WIMP, see, e.g.\cite{Roszkowski:2017nbc,Plehn:2017fdg}
for the latest reviews), direct, indirect and collider searches have so far failed to give an
uncontroversial signal of such particles. Among the alternative ideas for dark matter that have
emerged over the years, sub-GeV dark matter is gaining momentum, thanks both to a rich
upcoming experimental program and to the fact that, similarly to the WIMP, it relies on the robust,
UV-insensitive, thermal freeze-out mechanism to achieve the correct relic density (see
\cite{Battaglieri:2017aum} and\cite{Alexander:2016aln} for reviews).  These dark
matter scenarios typically involve a dark matter candidate interacting with SM particles
through a light mediator.  In this article we shall focus on a specific
class of models where the mediator is a new
gauge boson, $\ap$, corresponding to a spontaneously broken new abelian gauge group
$\udark$, because of their viability in providing a light thermal dark
matter as well as because of many experimental searches devoted to such models.  We will
refer to this new gauge boson
as the \textit{dark photon} in the following.

Since the new $\udark$ gauge group can mix with the
Standard Model $U(1)_Y$ gauge group, the dark photon acts as a proper mediator
between the dark and visible sectors. Such dark gauge groups have been particularly
often used in a dark matter context due to their interesting properties and experimental prospects for detection (some very recent examples are, e.g.\cite{Chu:2016pew,Correia:2016xcs,Knapen:2017xzo,Kuflik:2017iqs,Feng:2017drg,Duch:2017khv,Feng:2017vli,Wojtsekhowski:2017ijn,Feng:2017uoz}).
For instance, they can give rise to simple Self-Interacting Dark Matter models (SIDM, see\cite{Tulin:2017ara} for the latest
review) which could lead to better agreement between numerical simulations and the
astrophysical observations.

One of the simplest and experimentally-motivated way to generate the dark photon mass
perturbatively is through a ``dark Higgs mechanism''. This assumes the presence of an
additional dark Higgs boson which gives the dark photon its mass through a Vacuum
Expectation Value (VEV) $v_S$. Thus, a complete, self-consistent ``dark sector'' contains
a dark matter candidate, the dark photon and the dark Higgs boson. 
Crucially, both the dark Higgs boson mass and the dark photon mass are proportional to
$v_S$, so that a light dark photon should typically be accompanied by a light dark Higgs
boson.\footnote{The dark Higgs boson suffers from the same, and actually much larger naturalness problem
	as the Standard Model Higgs boson. We will assume that this problem is decoupled
	from our analysis (for instance that any supersymmetry-related fields are heavy
	enough to have a negligible influence). See in particular\cite{morrissey} for a
discussion of a dark sector in a supersymmetric context. } 
Note that a popular alternative for $U(1)$ extensions of the Standard Model consists in
introducing a Stueckelberg field along with a mass term for the new gauge boson, see
e.g.\cite{Feldman:2007wj}. We focus instead in this paper on the phenomenologically-richer
(in particular with respect to the pseudo-Dirac dark matter case) and experimentally
well-grounded Higgs mechanism.

Paradoxically, most of the literature on the field either focused on the dark Higgs boson,
with or without the dark photon, or assumed that it decouples from the rest of the
spectrum and concentrated on the dark matter and the dark photon only (one of the recent
exceptions is\cite{Choi:2016tkj} with a focus on the relic density constraint). In
contrast, we present in this paper two minimal, self-consistent and perturbative models
for the dark sector and systematically study the large part of the parameter space where
the dark Higgs boson is light. In this case the dark photon, dark matter and dark Higgs
boson must all be considered simultaneously. 

As we will see below, the most important characteristic of a light dark Higgs boson is the fact that its lifetime is typically of order of one second or longer.  Indeed, when the dark Higgs boson is lighter than
the dark photon and of twice the dark matter mass, its decay is particularly suppressed as it can only proceed
through a loop-induced coupling to light Standard Model particles. Such long-lived dark Higgs boson have been studied independently for several years and have been shown to possibly leave a signal in long baseline neutrino experiments and more generally in
so-called ``beam-dump'' experiments (see,
e.g.\cite{Batell:2009di,Bjorken:2009mm,Schuster:2009au,Essig:2009nc}). Light dark Higgs boson
originating for instance from the decay of a light meson can travel through the shielding of these
beam-dump experiments and subsequently decay in the downstream detector.\footnote{This is similar to
the idea that beam-dump experiments can create a detectable ``dark matter beam'' when dark
matter is light (typically below a few GeV) which has received more attention in recent
years (see,e.g.\cite{Batell:2009di,Batell:2014mga,Izaguirre:2015yja,deNiverville:2016rqh,Battaglieri:2016ggd}).}  We will re-evaluate this particular search strategy for detecting dark Higgs bosons and show that they are currently not sensitive enough to reach the thermal value target in our two minimal models.

The second main result is that the relic density calculation is thoroughly modified by the presence of new dark
matter annihilation channels involving a dark Higgs boson. In particular, the long lifetime of the
dark Higgs boson implies that the thermal freeze-out mechanism proceeds as in a two-component dark matter
scenario. However, its presence also opens up new additional $s$-wave annihilation channels for dark
matter at the time of recombination and leads therefore to severe bounds from CMB
observations\cite{Slatyer:2015jla,Liu:2016cnk}.

Finally, a long-lived
dark Higgs boson is constrained by Big Bang Nucleosynthesis (BBN) related
data\cite{Fradette:2017sdd,Kawasaki:2017bqm}, especially given that its metastable density obtained
from thermal freeze-out can be larger than that of the dark matter.  Nonetheless, we will show that
light dark Higgs bosons in our two minimal dark sector models have metastable density
substantially smaller than the  Higgs portal case and can
alleviate significantly the bounds presented in\cite{Fradette:2017sdd}.

The paper is organized as follows. We first present in Sec.~\ref{sec:models} two models of
the dark sector framework, as well as existing constraints on the dark photon
from various experiments. We then focus in Sec.~\ref{sec:dm} on the dark matter candidate
and the effect of the presence of the dark Higgs boson on its relic density and on the
constraints from CMB.  Section~\ref{sec:BD} discusses detection prospect for the dark
Higgs boson  in beam-dump experiments as well as constraints related to BBN.  Finally, in
Sec.~\ref{sec:res} we summarize our results and conclude.  The appendix contains additional details
on the calculation of dark Higgs boson production cross section from light mesons decay.

\section{Minimal dark sector models and bounds on the dark photon}
\label{sec:models}
We present in this section two minimal, self-consistent dark sector models for a sub-GeV
dark matter. As was discussed in the Introduction, such dark sectors typically include
three types of fields:
\begin{itemize}
 \item an extra gauge boson (called ``dark photon'' in the following) $\ap$ corresponding
	 to ``dark'' gauge group $U(1)_D$ with a gauge constant $\gzp$;
 \item a complex scalar $S$  with charge $q_S$, called henceforth ``dark Higgs boson''. It
	 spontaneously breaks the dark gauge group through a VEV,
	 $v_S$;
 \item a dark matter particle $\chi$ with charge $q_\chi$. We will consider both a complex
	 scalar and a Majorana fermion dark matter candidate. As usual, we will assume
	 that a discrete $\mathbb{Z}_2$ symmetry protects the dark matter from decaying.
\end{itemize}

\subsection{Lagrangian, masses and lifetimes}

The gauge and matter content that we are considering implies that the dark sector can be coupled to
the SM either through kinetic mixing between the two abelian gauge groups or by mixing between the SM Higgs $H$
and the dark Higgs boson $S$. While both portals are a priori open, in this article we will focus on the
vector portal. We will furthermore argue below that this is the most natural choice given the
sub-GeV mass domain we are interested in. The kinetic mixing can in principle arise from loops of heavy fields charged under both gauge groups. We will
assume in the following that they are safely decoupled at the energy scale that we consider.

Given that the dark matter candidates must be charged under the new gauge group $U(1)_D$,
care must be taken when choosing them such that the dark gauge group remains anomaly-free. In particular, this excludes one
single Majorana dark matter candidate, albeit a non-minimal scenario with a second heavier Majorana field canceling the anomaly is still possible. Consequently, we will consider in this paper two minimal, self-consistent, models for the dark
matter candidates:
\begin{itemize}
 \item model \pdf: the pseudo-Dirac fermion case, where a Dirac fermion $\chi = (\chi_L,\chi_R^\dagger)$ dark matter acquires additional Majorana masses from its Yukawa interactions with the dark Higgs boson;
 \item model \cs: the complex scalar dark matter case, where we also denote the dark matter
	 field by $\chi$.
\end{itemize}
The simplest charge assignment in the \pdf~case is a $U(1)_D$ charge
$+2$ for the dark Higgs boson $S$ and $\pm 1$ for the two dark matter fermions $\chi_L$ and
$\chi_R$ . In the \cs~case, we assign a charge $+1$ to the dark Higgs boson
$S$ and $+1$ to the complex scalar dark matter $\chi$. 

The effective Lagrangian for the dark photon vector and the dark Higgs boson fields in these two minimal dark sector
models is then given by
\begin{align}
	\mathcal{L}_{\ap} &= -\frac{1}{4}F^{\prime\mu\nu}F^{\prime}_{\mu\nu}
	-\frac{1}{2}\frac{\eps}{\cos\theta_w}B_{\mu\nu}F^{\prime\mu\nu} \ , \\ 
	\mathcal{L}_S&=(D^\mu S )^*(D_\mu S)+ \mu_S^2 |S|^2 - \frac{\lambda_S}{2} |S|^4 - \frac{\lambda_{SH}}{2} |S|^2  |H|^2 \ , 
	\label{lma}
\end{align}
while the DM field is introduced either as a scalar or a fermion through the Lagrangian
\begin{align}
	\mathcal{L}^{\rm DM}_{\pdf}&=\bar{\chi}\left( i\slashed{D}-m_{\chi}\right)\chi +
	V^{m}_{\pdf} (S,\chi) \ ,\\
	\mathcal{L}^{\rm DM}_{\cs}&=(D^\mu \chi)^*(D_\mu \chi)-m_\chi|\chi|^2 + V^{m}_{\cs} (S,\chi)
	\label{ldm} \ ,
\end{align}
where $V_{\pdf}$ and $V_{\cs}$ describe the mixing of the DM particle with the dark Higgs boson $S$. We parametrize them as
\begin{align}
 V_{\pdf} &= y_{SL} S \chi_L \chi_L + y_{SR} S \chi_R^c \chi_R^c + \textrm{ h.c.} \ ,   \\
V_{\cs} &= \lambda_{\chi} |\chi|^4 + \lambda_{\chi S} |\chi|^2 |S|^2 + \lambda_{\chi
 H} |\chi|^2 |H|^2 \ .
\end{align}

If $\mu_S^2 >  0$, and in the relevant limit where $\lambda_{SH} \ll \lambda_S, \lambda_H$, we can solve the tadpole equations for the VEVs of the SM Higgs $v_H$ and of the dark Higgs boson $v_S$, leading to
\begin{align}
 v_S^2 &= \frac{1}{\lambda_S} ( \mu_S^2 - \frac{\lambda_{SH}}{2 \lambda_H} \mu_H^2) \ ,\\ 
 v_H^2 &\simeq \frac{\mu_H^2 }{\lambda_H} \nn
\end{align}
where $\mu_H^2$ and $\lambda_H$ are respectively the SM Higgs mass term and self quartic coupling. At zeroth order in $v_S / v_H$, the dark Higgs boson mass $\ms$ and dark photon mass $\mzp$ are
\begin{align}
\label{eq:mass}
 \ms &= \sqrt{2 \lambda_S} v_S \ , \\
 \mzp &= \gzp q_S v_S = \left( \frac{q_S \gzp}{\sqrt{2 \lambda_S}} \right) \ms\ ,
\end{align}
where we have introduced the dark Higgs boson $\udark$ charge $q_S$. In particular, the dark Higgs boson is lighter than the dark photon when
\begin{align*}
\sqrt{2 \lambda_S} < q_S \gzp \ .
\end{align*}
This case will be of particular interest since the dark Higgs boson is then long-lived, as we will see in the next section. 

Notice that for typical SM-like values $\lambda \sim 0.1$ and $\gzp \sim 0.5$, the dark
Higgs boson is indeed lighter than the dark photon. Furthermore, when the dark gauge
coupling is chosen near its perturbativity bound with $\azp \equiv \gzp^2 / 4 \pi$ of
order $0.5$, then having a dark Higgs boson heavier than the dark photon leads to
$\lambda_S > 1.25 q_S^2$ and therefore possible non-perturbative behavior in the dark sector. For
large values of $\azp$, assuming the dark Higgs boson to be heavy enough to completely decouple from
the rest of the dark sector is hence impossible in a minimal perturbative
setup.

The kinetic mixing parameter should be small enough to avoid various experimental bounds discussed in the following sections. In a Grand Unified Theory context, the required small values for $\eps$  could be obtained from  loops of heavy
particles charged under both the SM hypercharge $U(1)_Y$ and the new $\udark$ gauge
group\cite{Holdom:1985ag}, with values between $10^{-2}$ and $10^{-5}$ depending on whether
the mixing is generated at one or two-loops.\footnote{For more details about the limit case of an almost decoupled dark sector with freeze-in realization of the correct relic density, see\cite{Berger:2016vxi}.} Notice that after diagonalizing the gauge kinetic terms, dark sector particles remain neutral under electromagnetism, but Standard Model fields acquire an $\epsilon$-suppressed coupling to the dark photon.

Finally, in the \pdf~case, the dark Higgs boson VEV leads to Majorana mass terms for the
left-handed and right-handed components of $\chi$. After diagonalizing the mass matrix, the lightest
eigenstate $\chi_1$ becomes our dark matter candidate. Notice that in principle $y_{SL} \neq y_{SR}$
so that gauge coupling of schematic form $\chi_1 \chi_1 V$ and $\chi_2 \chi_2 V$ are a priori
generated (albeit typically suppressed compared to $\chi_1 \chi_2 V$ term).
\subsection{Dark Higgs boson lifetime}
\label{sec:lifetime}
\begin{figure}
	\centering
	\subfloat[]{%
\raisebox{0.5\height}{\includegraphics[width=0.4\textwidth]{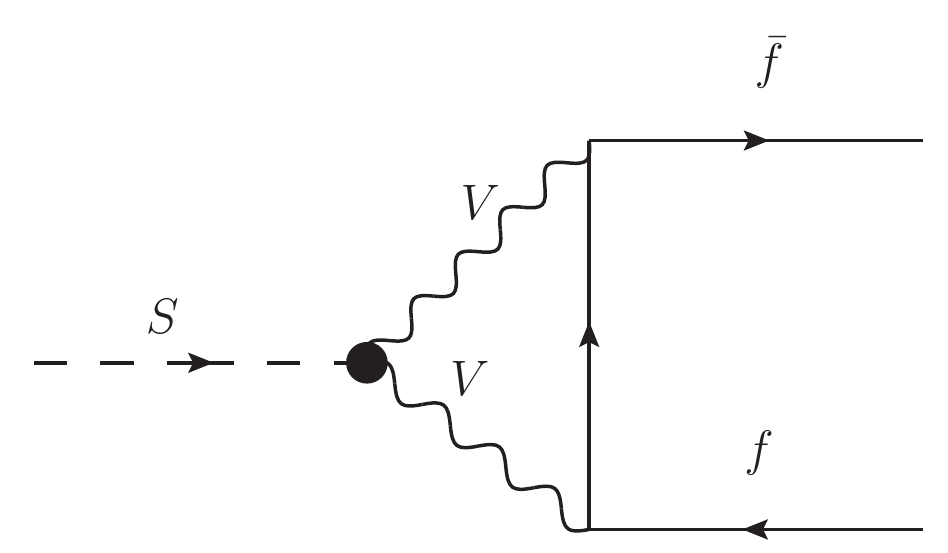}}
\label{fig:tauA}}
	\subfloat[]{%
\includegraphics[width=0.55\textwidth]{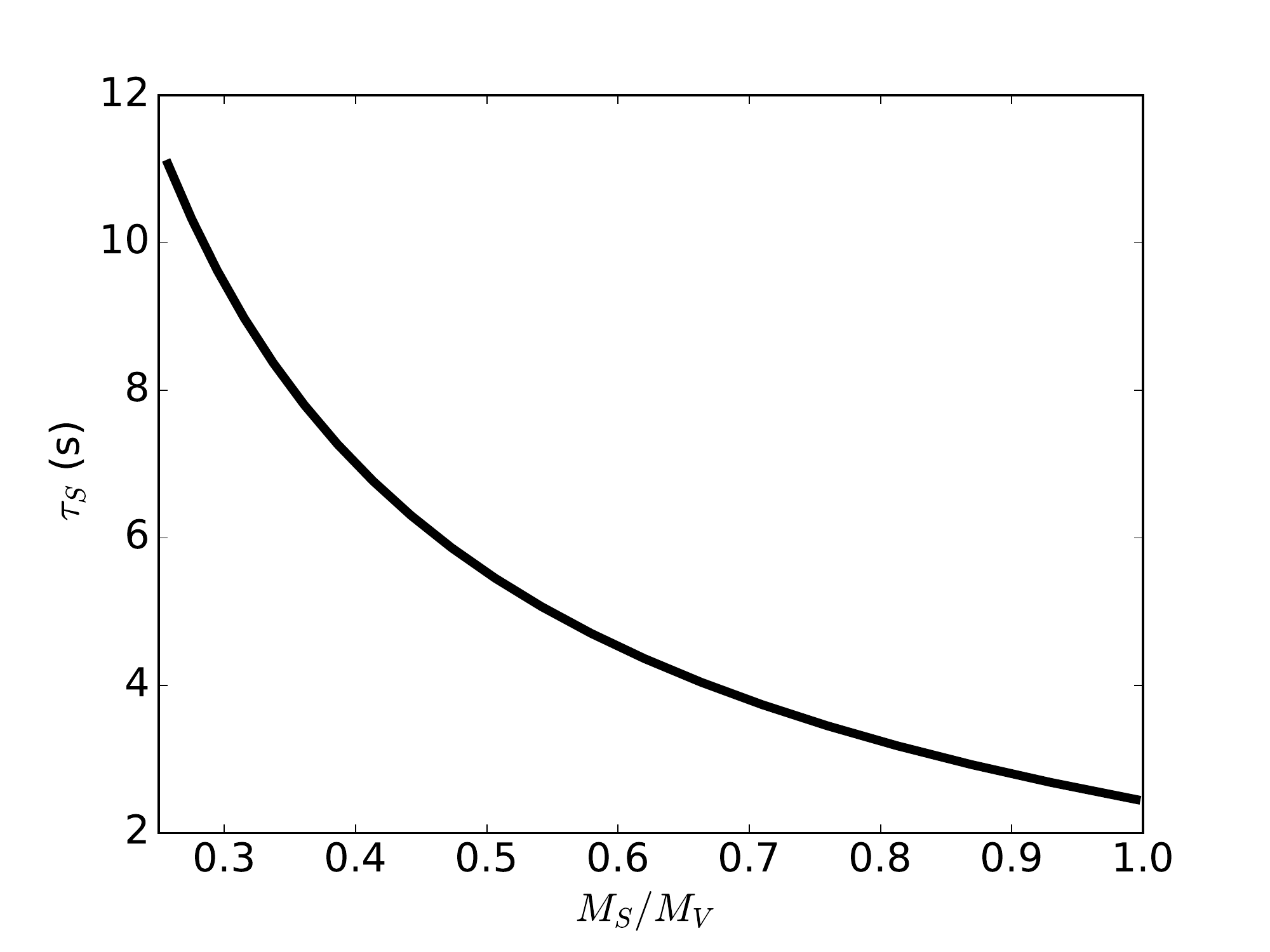}\label{fig:tauB}}
\caption{\a~Loop diagram for the dark Higgs boson decay with $f$ denoting an SM fermion whose coupling to $V$ is
	$\eps$-suppressed. \b~Dark Higgs lifetime in seconds as a function of the
	ratio $\ms/\mzp$ for $\azp = \aem$, $\eps = 0.001$ and $\mzp = 200$ MeV.} 

\end{figure}

When the tree-level decay of dark Higgs boson to dark matter is kinematically forbidden and
its mixing with SM
Higgs boson is negligible, the only decay mode available is through a triangular diagram of the form given in
Fig.~\ref{fig:tauA}.   Furthermore, when $\ms<2m_\mu$ the dominant decay mode
is $S \rightarrow e^+ e^-$ with
the dark Higgs boson width given by\cite{Batell:2009yf}
\begin{align}
 \Gamma_{S\rightarrow ee} = \frac{\azp \alpha^2 \eps^4 \ms}{2\pi^2} \frac{m_e^2}{\mzp^2} \left( 1-
 4\frac{m_e^2}{\ms^2} \right)^{3/2} \left|I(\frac{\ms^2}{\mzp^2},\frac{m_e^2}{\mzp^2}) \right|^2 \ ,
\end{align}
where the loop function is expressed as
\begin{align*}
 I(x_s,x_e) = \int_0^1 dy \int_0^{1-y} dz \frac{2-(y+z)}{(y+z)+(1-y-z)^2  x_e - yz x_s} \ .
\end{align*}
The above expressions also apply to the decay to muons by just replacing $m_e$ with $m_\mu$. In
particular, in the limit $m_e \ll \ms,\mzp$, we have $$ \Gamma_{S\rightarrow ee} \propto  \frac{\eps^4
m_e^2}{\mzp} \left(\frac{\ms}{\mzp} \right) \ .$$ The corresponding lifetime $\tau_S$ is presented in Fig.~\ref{fig:tauB} as a
function of $\ms/\mzp$, from which one can recover the exact value for any set of parameters using
the previous scaling relations.

As an order of magnitude estimate, we then have
\begin{align}
  \tau_S \propto 2 \cdot 10^{-3} \textrm{ s} \times \left( \frac{\aem}{q_S^2  \azp}\right) \left(
 \frac{ 10^{-3}}{\eps}\right)^4 \left( \frac{100 \textrm{ MeV}}{\ms}\right) \left(
 \frac{\mzp}{2 m_f}\right)^2 \ ,
\end{align}
where $f$ are the kinematically accessible SM fermions, $\aem$ is the
electromagnetic fine-structure constant. In particular for $\ms$ below the dimuon mass threshold
we find
\begin{align}
\label{eq:Slifetime}
\tau_S \propto 10 \textrm{ s} \times \left( \frac{\aem}{q_S^2  \azp}\right) \left(  \frac{ 10^{-3}}{\eps}\right)^4 \left( \frac{50 \textrm{ MeV}}{\ms}\right) \left( \frac{\mzp}{ 100 \textrm{ MeV}}\right)^2 \ .
\end{align}

In principle, the mixing between the Standard Model Higgs and the dark Higgs boson through
the mixing quartic coupling $\lambda_{SH}$ could lead to additional decay channels.
However, since the Higgs boson VEV contributes at tree level to the dark Higgs boson mass
by
$\lambda_{SH} v^2$ we need
\begin{align}
 \lambda_{SH}  \sim \frac{\ms^2}{v_H^2} \sim 10^{-8} \textrm{ - }  10^{-6} \ ,
\end{align}
for a dark Higgs boson mass between $10$ and $100$ MeV.
If the dark Higgs boson could only decay to $\epem$ through its mixing with the SM Higgs, its
lifetime $\tau_{{}_{S, H \textrm{mix}}}$ would then be parametrically given by
\begin{align}
\tau_{{}_{S, H \textrm{mix}}} \propto 1 \cdot 10^6 \textrm{ s} \times \left( \frac{100 \textrm{ MeV}}{\ms}\right)
\left( \frac{100 \textrm{ MeV}}{\mzp}\right)^2  \left( \frac{10^{-6}}{
	\lambda_{SH}}\right)^2 \left( \frac{q_S^2 \azp}{ \aem}\right)  \ .
\end{align}
This implies that, unless one is prepared to significantly tune the theory to ensure a
light dark sector while keeping a large $\lambda_{SH}$, the decay through SM Higgs mixing should be
significantly smaller than the loop-induced one.

In the rest of this article, we will therefore neglect the Higgs-portal related effects (which includes the quartic
$\lambda_{SH}$, but also for simplicity, the dark matter/Higgs quartic $\lambda_{\chi H}$
in the \cs~case).

\subsection{Constraints on the dark photon}
\label{sec:otherconst}
If the dark matter is heavier than half of the dark photon mass, the dark photon decays mainly into a pair of leptons. This minimal scenario is
mostly constrained by searches for bumps in the dilepton invariant spectrum at
NA-48/2\cite{Batley:2015lha}, BaBar\cite{Lees:2014xha} and LHCb\cite{Aaij:2017rft}, setting bounds for $\eps \lesssim 10^{-3}$. Slightly less competitive
bounds also arise from rare meson decays. 
For very small kinetic couplings leading
to a long-lived dark photon decaying to visible sector, one can also obtain bounds from
electron beam-dump experiments like E137, E141 or E774. These searches hence give a lower bound
on the kinetic mixing for a dark photon with mass in the tens of MeV range (see,
e.g.\cite{Alexander:2016aln} for a summary of the current bounds).

The most relevant case for the parameter space considered here is when the dark photon decay channel
to dark matter is kinematically open,
so that one should search for the missing momentum carried away by the Dark matter particles\cite{Andreas:2013lya,Izaguirre:2014bca}.
The strongest
bounds are currently set by searches at BaBar\cite{Lees:2017lec} and NA64\cite{Banerjee:2017hhz}. 
More
precisely, the BaBar analysis searches for narrow peaks in the distribution of missing mass arising
from $e^+ e^- \rightarrow \gamma V $ events. Their limit excludes the region $\varepsilon>10^{-3}$ for the dark photon mass range we consider,
which in particular rules out the dark photon explanation for the $(g-2)_\mu$ excess. Secondly, the
NA64 Collaboration recently released bounds on the decay $V \rightarrow \textrm{invisible}$. Their limits significantly exceed the one set by
BaBar for $\mzp \lesssim 100 $ MeV, reaching $\eps<10^{-4}$ below $10 $ MeV. An explicit visualization of
these bounds will be shown below in Fig.~\ref{fig:NA64} in Sec.~\ref{sec:CMB}. Note that the projected bounds from the LDMX
proposal (see, e.g.\cite{Battaglieri:2017aum}) will cover almost all of the parameter space consistent with
the relic density thermal value target as shown in Fig.~\ref{fig:NA64}.

\vspace{1cm}

\begin{table}[t!]
\begin{center}
\begin{tabular}{|c|m{7cm}|c|c|}
\hline
\rule{0pt}{1.25em}
\textbf{Parameter} &  \textbf{Description} & \textbf{Range} &  \textbf{Prior}\\[0.15em]
\hline
\hline
\rule{0pt}{1.em}
$\lam_S$ & Dark Higgs boson quartic coupling & $\,10^{-4},\,0.25$ & Log \\
\hline
\rule{0pt}{1.em}
$\gzp$ & Dark gauge coupling & $\, 5 \cdot 10^{-3} ,\,1.25$ & Linear \\
\hline
\rule{0pt}{1.em}
$\mzp$ & Dark photon mass & $\,10,\,500$ & Log \\
\hline
\hline
\rule{0pt}{1.em}
$\eps$  & Kinetic mixing parameter & $\,1.5\cdot 10^{-5},\,1.5\cdot 10^{-3}$ &  Log \\
\hline
\hline
\rule{0pt}{1.em}
$m_\chi$  &  Complex scalar DM mass & $10,\,500$  (\cs) &  Log \\
  & Dirac DM mass & $10,\,500$ (\pdf) &  Log \\
\hline
\rule{0pt}{1.em}
$\lambda_{S\chi}$  &  Quartic mixing between the dark Higgs boson and DM & $\,10^{-3},\,0.2$ (\cs)&  Log \\
\hline
\rule{0pt}{1.em}
$\lambda_{\chi}$  &  Self DM quartic coupling& $\,5\cdot 10^{-4},\,0.1$ (\cs)&  Log \\
\hline
\rule{0pt}{1.em}
$y_{S L}$  &  Left-handed DM-dark Higgs boson Yukawa coupling  & $\,10^{-3},\,0.7$ (\pdf)&  Log \\
\hline
\rule{0pt}{1.em}
$y_{S R}$  &  Right-handed DM-dark Higgs boson Yukawa coupling  & $\,10^{-3},\,0.7$ (\pdf)&  Log \\
\hline
\end{tabular}
\caption{Parameters of the models analyzed in this work. All parameters are initialized at the electroweak scale.
Dimensionful quantities are given in MeV and $\mev^2$.}
\label{tab:models}
\end{center}
\end{table}%

In the following and for all the numerical results, we used the code
\texttt{MultiNest}\cite{Feroz:2008xx} to direct the scanning procedure, based on the dark matter
relic density. All data points presented in this paper are therefore compatible with the result from
the Planck Collaboration\cite{Ade:2015xua} $\abund = 0.1188 \pm 0.0010$ at $95 \% $ CL. The
interfaces with the various public codes used here is done with the help of the private code BayesFITS. We use a
slightly modified version of \micromegas~v.4.3.5\cite{Belanger:2014vza} (and of its
two-component dark matter module). We evaluate the spectrum from the non-SUSY
SPheno\cite{Porod:2003um,Porod:2011nf} code generated by 
\SARAH~(see Refs.\cite{staub_sarah_2008,Staub:2012pb,Staub:2013tta}). We use renormalization group
evolution of the hidden sector parameters to ensure their perturbativity up to the electroweak
scale, and evaluate all masses at tree-level due to the light scale considered. Finally, the
estimation of the number of events in beam-dump experiments is obtained from a substantially
modified version of BdNMC from\cite{deNiverville:2016rqh} (more particularly, we have used the
original code to extract the distributions of initial mesons and expanded its routines to the
production and detection processes relevant for the dark Higgs boson).

In the following, we will restrict our analysis to the case where the dark Higgs boson is below the
dimuon threshold, so that it can only decay to an $e^+e^-$ pair. The dark photon is
also considered to be lighter than around $500$ MeV, so that the leptonic decay channels still
dominate its decay width compared to hadronic ones (see\cite{Liu:2014cma}). We summarize the independent parameters and their scanned ranges and priors in Table~\ref{tab:models}. Note that we do not vary the SM Higgs parameters. In particular, we take advantage of the
relation~\eqref{eq:mass} to trade $v_S$ for $\mzp$ as an input parameter, so that we vary $\gzp,
\eps, \ms, \mzp, m_\chi, y_{S L}$ and $ y_{S R}$ in the \pdf~model and $\gzp, \eps, \ms, \mzp,
m_\chi, \lambda_{S \chi}$ and $ \lambda_{\chi}$ in the \cs~model.

\section{Light DM phenomenology}
\label{sec:dm}

In this section we discuss the phenomenology of the light DM candidate in our two minimal dark sector
models.  We focus particularly on the relic density constraints and on the bounds from  CMB power spectrum for $s$-wave
annihilation processes occurring during the recombination era. We begin with a discussion of relic density for the pseudo-Dirac fermion (\pdf~case) and complex scalar
(\cs~case) DM candidates. In the following, the dark matter mass is denoted by $\mdm$, which hence
refers to the mass of lightest mass eigenstate $\chi_1$ in the \pdf~case.

\subsection{Relic density}

The relic density of DM in the standard freeze-out scenario is obtained by solving the
following Boltzmann equation
\begin{equation}
	\frac{dn_\chi}{dt}+3Hn_\chi=-\csv\left( n_\chi^2-(n_\chi^\mathrm{eq})^2 \right) \ ,
\end{equation}
where $n_\chi$ is the density of the DM species and $\csv$ is the thermally averaged
annihilation rate of DM.  The thermally averaged annihilation rate is given by\cite{Gondolo:1990dk}
\begin{equation}
	\langle\sigma v\rangle=\frac{1}{8 \mdm^4TK_2(\mdm/T)}\int_{4\mdm^2}^\infty
	\sigma\sqrt{s}\left( s-4\mdm^2 \right)K_1(\sqrt{s}/T)ds \ ,
	\label{}
\end{equation}
where $K_1$ and $K_2$ are modified Bessel's functions.  A useful parametrization of the
annihilation rate is in terms of $s$-wave and $p$-wave annihilations like $\csv~\equiv~\sigma_0
x^{-n}$, with $x=m_\chi/T$.  Here $n=0$ for $s$-wave and $n=1$ for $p$-wave annihilation.  In this
parametrization, $x$ at freeze-out is given by\cite{Kolb:1990vq}
\begin{align}\nonumber
	x_f=~&\mathrm{ln}\left( 0.038(n+1) \frac{g}{\sqrt{g_*}} M_{Pl} \mdm \sigma_0\right)\\
	&-\left(n+\frac{1}{2}\right)\mathrm{ln}\left[ \mathrm{ln}\left(
	0.038(n+1) \frac{g}{\sqrt{g_*}}M_{Pl} \mdm \sigma_0 \right) \right] \ ,
	\label{xf}
\end{align}
where, following the notation of\cite{Kolb:1990vq}, we note that $g$ represents the DM degrees of freedom, while $g_{*}$ and $g_{*,s}$
represent the number of relativistic degrees of freedom at freeze-out.
With the above expression for $x_f$ one can write the approximate expression for relic
density as
\begin{align}
	\Omega
	h^2=0.1 \left(\frac{(n+1)x_f^{n+1}}{(g_{*s}/g_*^{1/2})}\right)\frac{ 10^{-26} \textrm{ cm}^3/s}{\sigma_0} \ .
	\label{oh2}
\end{align}

In the two minimal dark sector scenarios we consider, the dark matter particle can be either a pseudo-Dirac fermion (\pdf~case) or a
complex scalar (\cs~case). In both cases, including the dark Higgs boson field leads to
several new annihilation channels in a similar manner to the usual supersymmetric WIMP. The usual
behavior considered by the previous literature corresponds to the case when the dark Higgs boson is
significantly heavier than the dark matter candidate so that annihilation into dark Higgs boson is
suppressed even with thermal effects included. The dominant process is a $s$-channel annihilation
to SM particles through an off-shell dark photon with the annihilation cross-section, for instance in the~\cs~case given by\cite{Boehm:2003hm,deNiverville:2011it}
\begin{align}
 \sigma_0 = 2.8 \cdot 10^{-25} \textrm{cm}^3/\textrm{s}   \times \left(\frac{\eps}{10^{-3}}\right)^2\left(\frac{\azp}{\aem}\right)
	\left(\frac{\mdm}{100\:\mathrm{MeV}}\right)^2\left(\frac{100\:\mathrm{MeV}}{\mzp}\right)^4 \ .
\end{align}
Using the above expression for $\sigma_0$ in Eq.~\eqref{xf} and Eq.~\eqref{oh2} we arrive at
the following estimate for the relic density
\begin{equation}
	\Omega h^2\sim 0.1 \times \left(\frac{10^{-3}}{\eps}\right)^2\left(\frac{0.1}{\azp}\right)
	\left(\frac{25\:\mathrm{MeV}}{\mdm}\right)^2\left(\frac{\mzp}{75\:\mathrm{MeV}}\right)^4 \ .
	\label{oh2est}
\end{equation}

On the other hand, for $\mdm\sim\ms$, dark matter annihilation into final states involving dark
Higgs boson become relevant. They proceed either through a $t$-channel exchange of a dark matter particle,
or through a dark Higgs boson $s$-channel.  These new mechanisms alone could explain the current relic
density for a dark sector coupling between dark Higgs boson and dark matter in the range we
consider, and therefore have to be taken into account. A key complication of this setup is that the
dark Higgs boson is a metastable particle with lifetime above $0.01\,$s in almost all of our parameter
space.  Consequently, thermal freeze-out proceeds akin to a two-component dark matter scenario.
This is especially relevant when the mass of the dark matter and of the dark Higgs boson are of the
same order, so that both $\chi \chi \rightarrow S S $ and $S S \rightarrow \chi \chi $ processes are
occurring at the time of dark matter freeze-out.
This annihilation channel is similar to the ``secluded'' regime in classic Higgs-portal scenarios\cite{Pospelov:2007mp,Kouvaris:2014uoa,Krnjaic:2015mbs} although the metastability of the dark Higgs boson implies in our case that the reverse processes $S S \rightarrow \chi \chi $ must be included compared to these references.
Furthermore, in the case of the \pdf~ model, the fact that we consider the Yukawa couplings to the two Weyl components to be different in general (i.e, $y_{SL} \neq y_{SR}$ in
contrast with\cite{Izaguirre:2017bqb}) implies that the annihilation channels
$\chi_{_1}\chi_{_1}\to e^+e^-$ and $\chi_{_1}\chi_{_2}\to SS$ are also available.

\begin{figure}
	\centering
\subfloat[]{\includegraphics[width=0.49\textwidth]{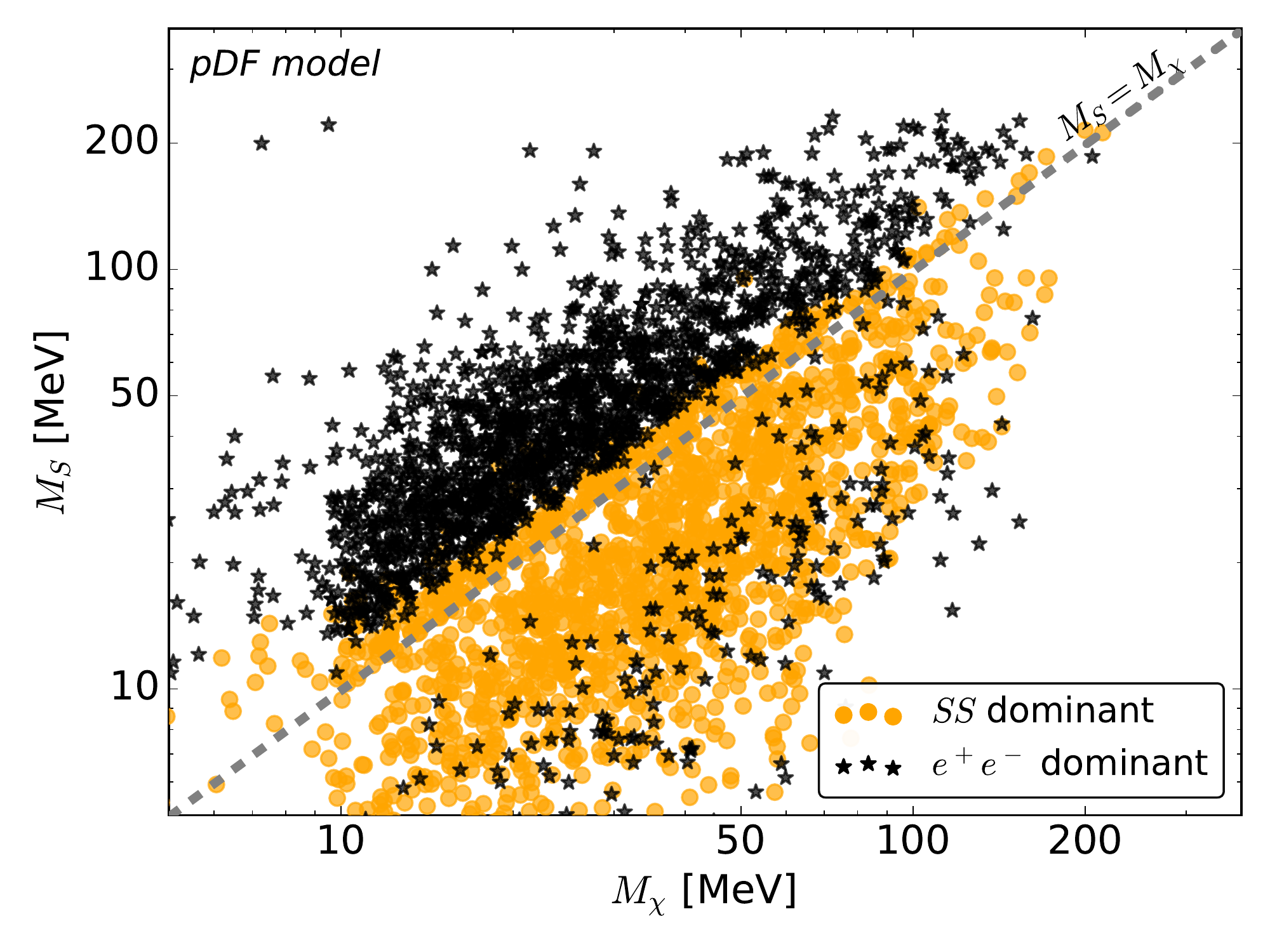}}
\subfloat[]{\includegraphics[width=0.49\textwidth]{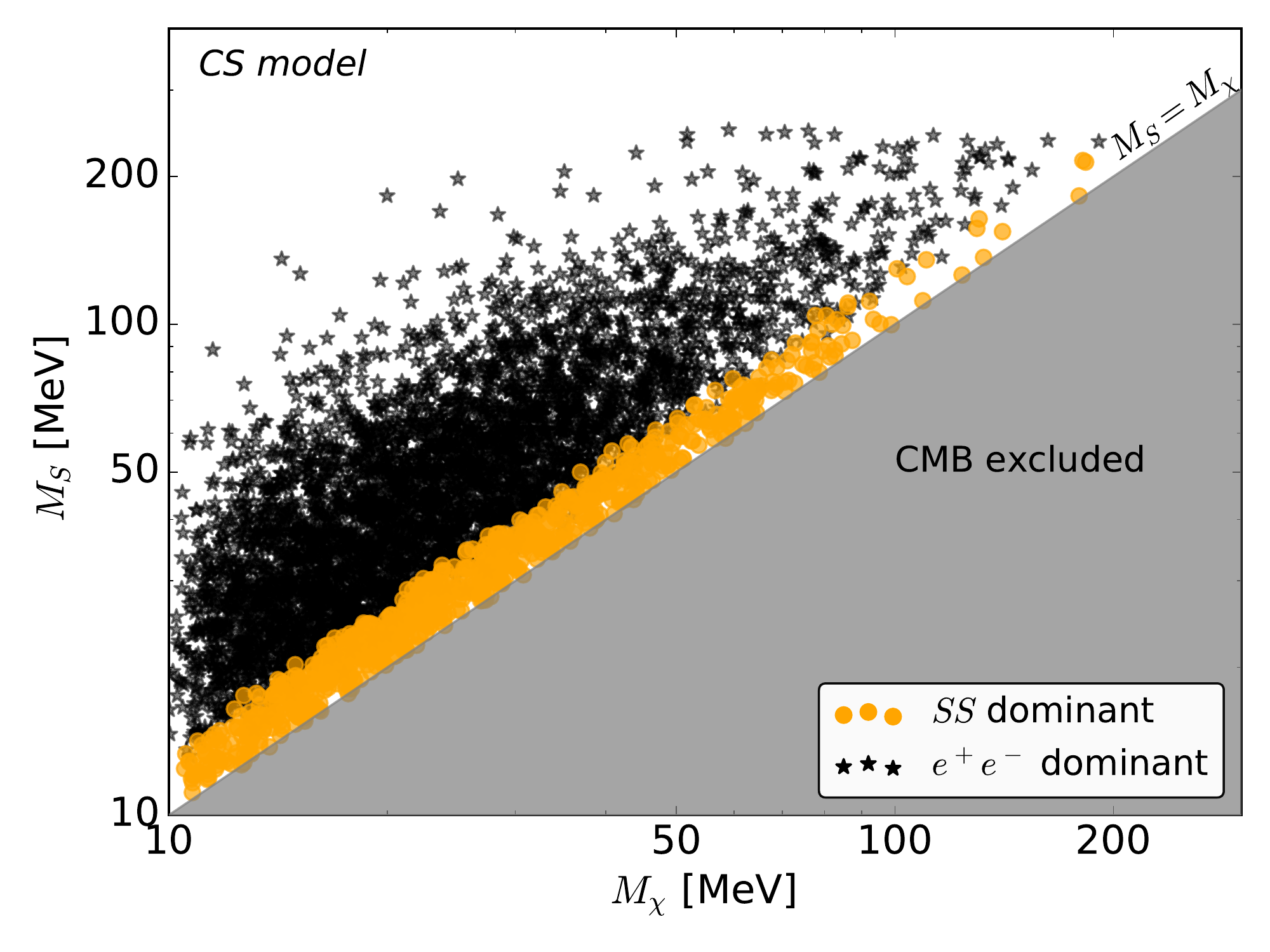}}
\caption{Points satisfying the dark matter relic density constraints in the $\mdm - \ms$ plane,
sorted according to the dominant annihilation channels at freeze-out in the \pdf~case \a~and the \cs~case \b. In \b, the region with $\mdm  > \ms$, which is excluded by CMB bounds, has been indicated.}
\label{fig:relic}
\end{figure}

In Fig.~\ref{fig:relic} we represent the relevant annihilation channels that contribute to
the relic density in the \cs~case and the \pdf~case.  We see from the figure that in the \cs~case,
when $\ms  \lesssim \mdm$ the $SS$ channel dominates (this region is excluded by CMB bounds) and when $\ms \gtrsim \mdm$ there are no $S$ final states with only the $e^+e^-$
channel remaining available to achieve the correct relic density.  For the \pdf~case the picture is more complicated due to the presence of
coannihilation channels.\footnote{ We have estimated the dominant annihilation cross-sections by summing the contributions from both annihilation and co-annihilation channels.} However, one still sees the $e^+e^-$ channel being predominant in the
$\ms>\mdm$ region, and the $SS$ channel dominating when $\mdm\simeq\ms$.  
In the very low mass region ($\mdm < 10$ MeV), our choice of parameter range (and in particular $m_\chi> 10$ MeV) implies that most of our points have a very large splitting between both dark matter components $\chi_1$ and $\chi_2$.\footnote{The mass matrix for dark matter in this case has a seesaw structure, which leads to the large splitting. There is no such mechanism for the \cs~case.} 
In particular, the channel $ \chi_1 \chi_1 \rightarrow e^+ e^-$ then becomes the dominant annihilation channel.
Finally, the presence of the ``reverse'' $S S \rightarrow \chi \chi $ channel implies that for some of our points in which the $\chi \chi \rightarrow S S $ annihilation process dominates, the thermal value target is in fact achieved by the subdominant channel $e^+e^-$, while  for some points in the $\mdm>\ms$ region the choice of couplings $y_{SL},\,y_{SR}$ and $\azp$, leads to $e^+e^-$ being the dominant channel.

Thus the presence of dark Higgs bosons changes significantly the relic density evaluation in our two
models. However, it also leads to two additional difficulties. First, the presence of a large metastable
density of dark Higgs boson after thermal freeze-out may lead to strong tensions with BBN.  We will explore
this aspect in Sec.~\ref{subsec:BBN}. Second, as we will see in the next section, the presence of
the new annihilation channels, while significantly reducing the constraints arising from the relic
density, may on the other hand be in strong tension with indirect bounds from CMB power
spectrum.

\subsection{Direct and indirect detection bounds}
\label{sec:CMB}
The CMB power spectrum has been measured with high precision and as such can impose stringent
constraints on the nature of DM.  In particular DM that injects energy in the form of
electromagnetically interacting SM particles in the inter-galactic medium (IGM) can significantly
alter the recombination history of the universe by ionizing and heating the IGM gas.  Such
injections from DM annihilation can be parametrized as $p_{ann}=f\csv/\mdm$, where $f$ denotes the
efficiency with which the energy injected by DM annihilations is transferred to the IGM.  Usually
the constraints from $s$-wave DM annihilations which do not depend on velocity of DM can be very
stringent and virtually rule out most models with $m_\chi<10$ GeV\cite{Slatyer:2015jla}.  Since electrons and photons are
the most efficient at ionizing the IGM, the annihilation channels that are most severely constrained
produce $e^-$s and $\gamma$-rays in their final states.


For the \pdf~model, when $\lambda_{S\chi_L} \neq \lambda_{S\chi_R}$ annihilation into
an $\epem$ pair as $\chi_1 \chi_1 \rightarrow V^* \rightarrow e^+ e^-$ becomes accessible.
It is however safely suppressed by mixing matrices elements and the off-shell nature of the $V$ in
all our parameter space.   



The situation is very different in the \cs~model, as $t$-channel annihilation into dark Higgs boson $\chi
\chi \rightarrow S S$ is completely unsuppressed when $\ms < \mdm$. Hence CMB bounds essentially
rule out this portion of the parameter space. 
Notice that in both cases, if $\mdm > \mzp$, other
annihilation channels involving the dark photon
open up which could lead to more severe bounds. However, they typically also significantly reduce the relic density, limiting the possibility to reach the thermal value target for the range of $\eps$ we consider in this paper. 

The bounds from CMB depend in principle on the annihilation
products (in particular they have been calculated for the $e^+e^-$, and $\epem$ via $S$ decays). However, they do not differ significantly in the dark matter mass
region we are interested in.
In the very low dark matter mass region of our plot ($\mdm \lesssim 10$ MeV), BBN-related bounds from energy injection from dark matter annihilation at freeze-out could become relevant\cite{Nollett:2013pwa}. However, they are model dependent and may in particular be modified due to the presence of a potentially long-lived dark Higgs. 


\begin{figure}
	\centering
\subfloat[]{\includegraphics[width=0.49\textwidth]{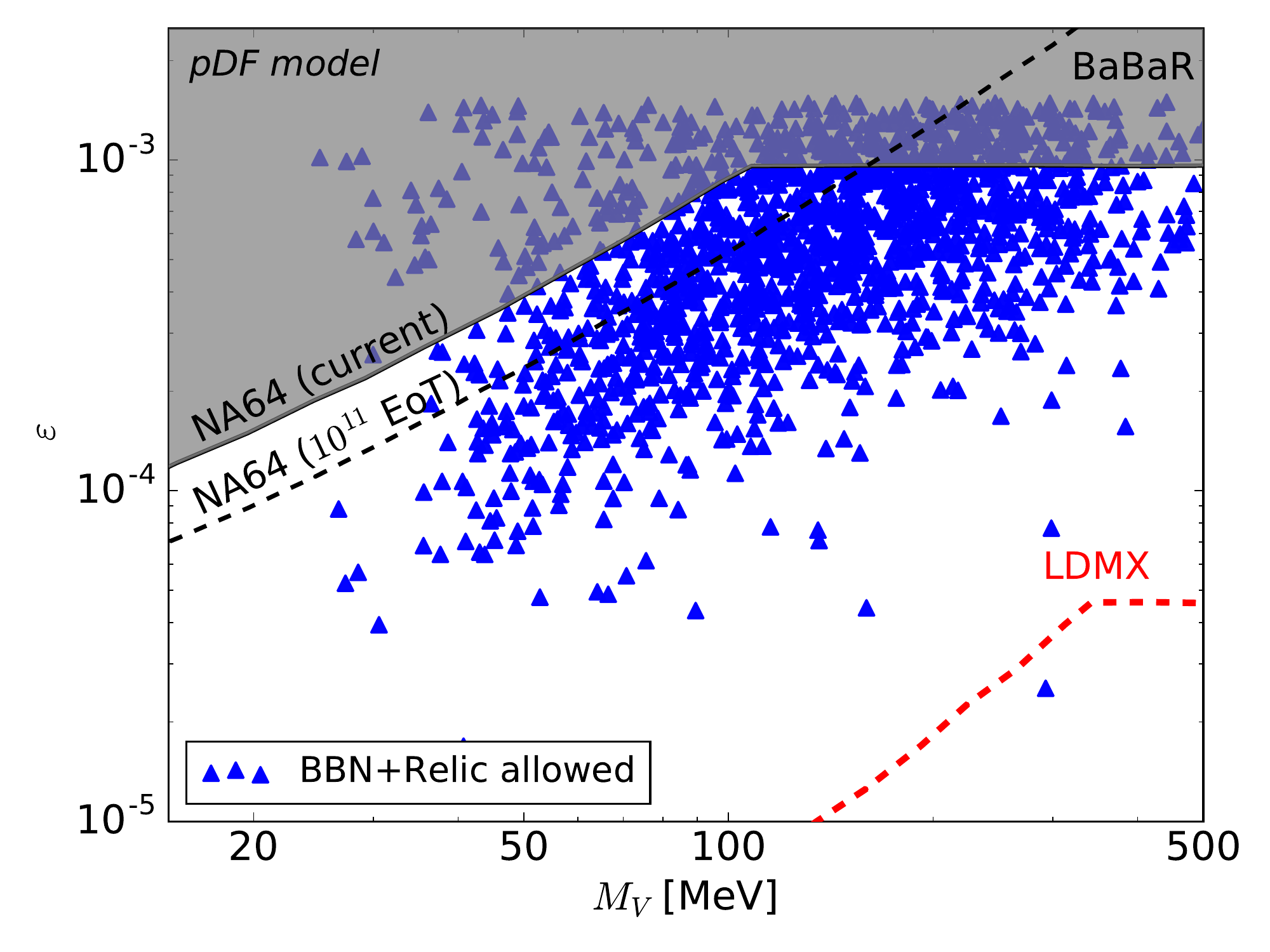}}
\subfloat[]{\includegraphics[width=0.49\textwidth]{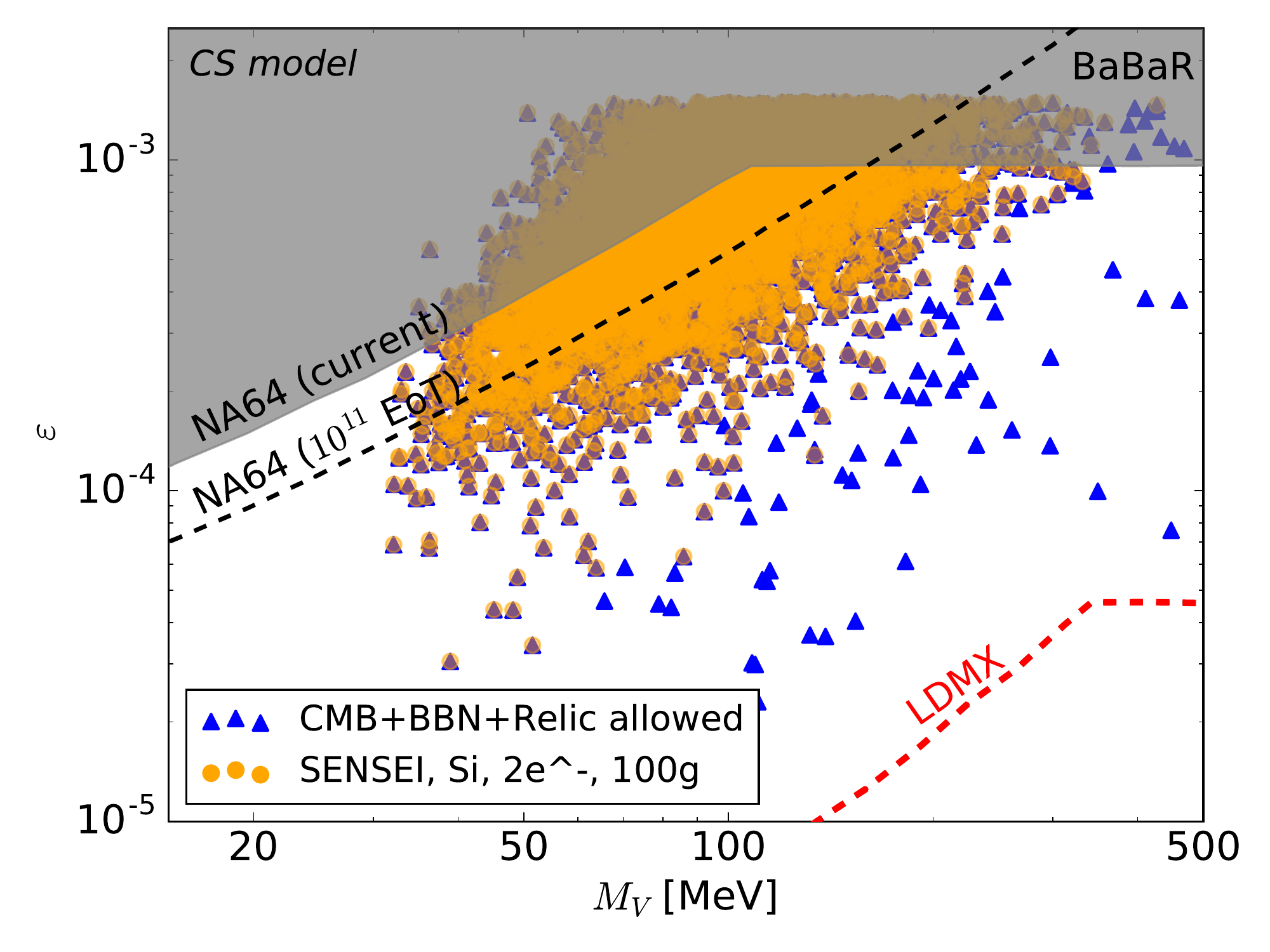}}
\caption{Constraints on the dark photon mass $\mzp$ and on the kinetic mixing parameter
$\eps$ from NA64 and BaBaR missing energy searches, with the projected bounds on a short 
time scale of SENSEI\cite{Tiffenberg:2017aac,Battaglieri:2017aum} with one year exposure 
of a $100$g detector (bounds from one year exposure of the annual modulation signals at 
Xenon1T according to\cite{Essig:2017kqs} are essentially similar) and from the full 
dataset of NA64 (corresponding to $10^{11}$ electrons on target, 
see\cite{Battaglieri:2017aum}). Points satisfying the dark matter relic density and 
relevant BBN and CMB constraints are shown for the \pdf~model \a~and for the \cs~case \b. 
The dashed red line represents the projected LDMX bound\cite{Battaglieri:2017aum}.}
\label{fig:NA64}
\end{figure}


Finally, in the case of complex scalar dark matter \cs, direct detection experiments
searching for DM scattering through electron recoil are also relevant for 
sub-GeV dark matter.  Different target materials such as noble liquids, semiconductors, 
scintillators and superconductors have been proposed for such searches (see 
\cite{Battaglieri:2017aum,Alexander:2016aln} for a discussion of these searches).  In the
case of noble liquid targets, searches for annual modulation signals through 
electron recoil were performed at
XENON10 and XENON100\cite{Essig:2012yx,Essig:2017kqs}, leading to the following bounds\cite{Kaplinghat:2013yxa}
\begin{align}
 \sigma_{\rm Xe}^{\rm SI} &~\simeq~ 4 \cdot 10^{-39} \textrm{cm}^2 \times   \left(  \frac{ \eps}{10^{-3}}\right)^2 \left( \frac{\azp}{ 0.01}\right) \left( \frac{100 \textrm{ MeV}}{\mzp}\right)^4 \nn \\
 &~\lesssim~ \scr{L}(\mdm)\cdot 10^{-38} \textrm{cm}^2 \ ,
\end{align}
where the last inequality is the derived XENON10/XENON100 bound $ \scr{L}(\mdm)$ which depends on the
precise dark matter mass (see\cite{Essig:2017kqs}).  In addition, experiments based on 
semiconductors using silicon CCDs like SENSEI\cite{Tiffenberg:2017aac,Battaglieri:2017aum} 
can also improve upon these bounds.  We present in 
Fig.~\ref{fig:NA64} the
corresponding bound from SENSEI as function of the dark matter mass for all points of 
our scans
satisfying the relic density constraint.  The projected bound from SENSEI can
probe almost all of the parameter space where we found the correct relic density (they are furthermore almost similar to one expected from annual modulation signals at XENON1T\cite{Essig:2017kqs}).  In 
future experiments using superconducting detectors based on aluminium can also probe 
this region of parameter space, but are perhaps more suited for sub-MeV range of masses. 
Finally, the rest of the parameter space will be totally covered by medium-term experiments, such as DAMIC-1K\cite{Battaglieri:2017aum}.


\section{Light dark Higgs boson}
\label{sec:BD}

We now turn to the second light state of our dark sector: the dark Higgs boson. As we have shown in Sec.~\ref{sec:lifetime}, this particle is long-lived in most of our parameter space. We explore in this section two consequences of this long lifetime: the detection prospects at proton beam-dump experiments, and the constraints from BBN-related observables.

\subsection{Beam dump experiments}
\label{subsec:bdump}

Fixed target experiments are well suited for the detection of light dark sector particles. They
typically involve a high-intensity, but relatively low-energy proton or electron beam
impacting the target, producing a shower of secondary particles, which are later disposed
off in a large shielding. Long-lived or stable dark matter particles are produced at a low
rate in the target, but since they interact very weakly with the shielding, they travel to
a downstream detector which can  subsequently detect them.

In particular, when the dark photon decays into dark matter particles, it effectively
produces a ``dark matter beam'' and the possible scattering of dark matter in the
detector can then be
estimated\cite{Batell:2009di,Batell:2014mga,Izaguirre:2015yja,deNiverville:2016rqh,Battaglieri:2016ggd}. In 
particular, a case comparable to our fermion dark matter scenario \pdf\ has been studied
in\cite{Izaguirre:2017bqb}.

In this section, we will focus instead on
examining the dark Higgs boson detection prospects in 
three proton beam-dump experiments:
LSND\cite{Athanassopoulos:1996ds}, miniBooNE\cite{AguilarArevalo:2008qa}  and the
proposed SBND experiment at Fermilab\cite{Antonello:2015lea}. The details of the
experimental setups are presented in Table~\ref{tab:exppars}. These three experiments rely on proton beams
with relatively low energy so that we expect dark sector production through
bremsstrahlung and direct production to be sub-dominant compared to the meson decay
mechanism\cite{deNiverville:2016rqh}.

Notice that past electron beam-dump experiments, like E137\cite{Riordan:1987aw}, can also lead to dark
sector beams through dark photon production by bremsstrahlung. However, the bounds on the kinetic
mixing parameter $\eps$ 
derived from dark Higgs boson production and decay at these facilities 
were found in\cite{morrissey} (in a context roughly similar
to ours -- albeit in a supersymmetric model) to be always significantly weaker than the current
missing energy bound $\eps < 10^{-3}$. The case studied in\cite{Izaguirre:2017bqb}, which we will
considered in more details at the end of this section, is a notable exception. 

\begin{table}[t]
\centering
\begin{center}
\makebox[\textwidth][c]{
\begin{tabular}{l| c c c c c}
Name & Energy  & Target Material & Distance & Length & Area \\
\hline
\rule{0pt}{3ex}LSND & $0.798$ GeV  &  Water/high-Z metal & $34$ m & $8.3$ m & $25.5$ m$^2$ \\
MiniBooNE & $8.89$ GeV  & CH$_2$ & $490$ m & Sphere &  $R=2.6$ m \\
SBND & $8.89$ GeV & CH$_2$ & $112$ m & $5$ m & $16$ m$^2$ \\
\end{tabular}
}
\caption{Summary of the relevant characteristics of the experiments considered. Detector distances are taken from the beam target to the center of the detector. LSND has a cylindrical geometry, MiniBooNE a spherical one and SBND should have a square intersection with the beam axis.}
\label{tab:exppars}
\end{center}
\end{table}

\begin{table}[t]
\centering
\begin{center}
\makebox[\textwidth][c]{
\begin{tabular}{l|c c  c c c }
Experiment & $\pi^0$ Distribution &$N_{\pi^0}$ & $N_{\eta} / N_{\pi^0}$ & $N_{\rho} / N_{\pi^0}$ &  $N_{\omega} / N_{\pi^0}$ \\
\hline
\rule{0pt}{3ex}LSND &Burman-Smith& $10^{22}$ & / & / & / \\
MiniBooNE &Sanford-Wang & $2 \cdot 10^{20}$ & 0.33 & 0.05 & 0.046 \\
SBND & Sanford-Wang & $6.6 \cdot 10^{20}$ & 0.33 & 0.05 & 0.046
\end{tabular}
}
\caption{Summary of the relevant characteristics of mesons productions in the experiments considered. Note that the lower energy at LSND prevents the production heavier mesons.}
\label{tab:expmeson}
\end{center}
\end{table}

\subsubsection{Dark Higgs boson production through meson decay}
\label{subsubsec:dhprod}

Proton beam-dump experiments could be practically seen as light meson factories, with
around one neutral pion created for each proton on the target. We furthermore include the
production of heavier $\eta$, $\rho$ and $\omega$ mesons. The relevant number of mesons produced
in each experiment is given in Table~\ref{tab:expmeson} based
on\cite{deNiverville:2011it,Batell:2014yra}. We simulate their kinematic distribution by using a weighted
Burman-Smith distribution to account for the different target material used by the LSND
experiment over its lifetime (water, then high-Z metal) and an averaged $\pi^+$ and
$\pi^-$ Sanford-Wang distribution for MiniBooNE and SBND.

\begin{figure}
	\centering
\subfloat[]{\includegraphics[width=0.45\textwidth]{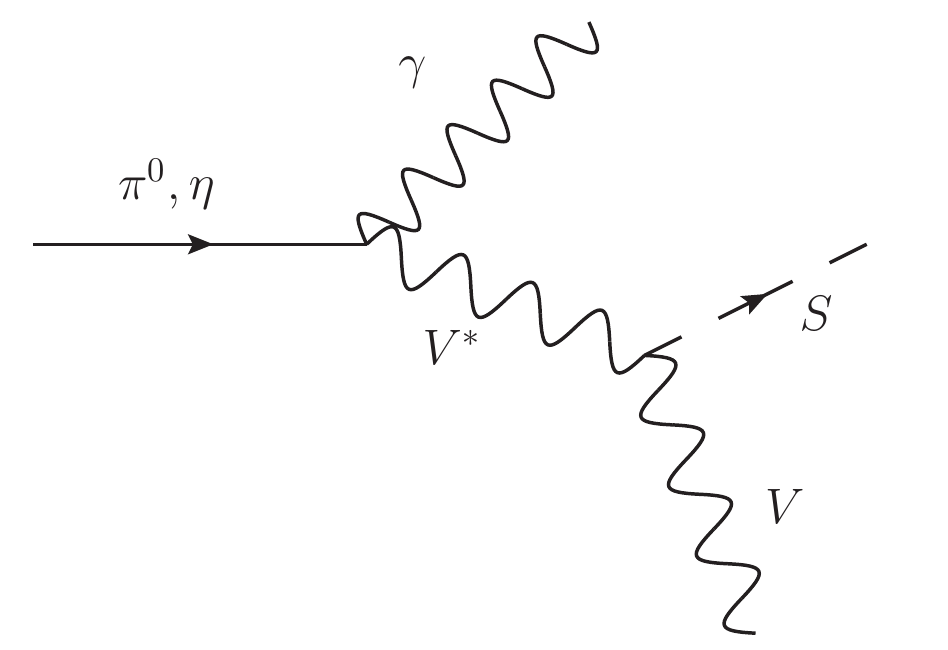}}
\subfloat[]{\raisebox{0.5\height}{\includegraphics[width=0.45\textwidth]{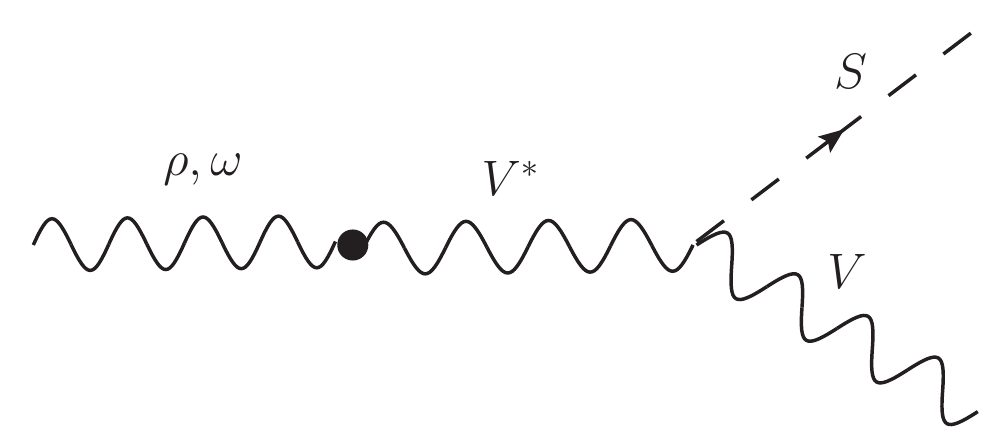}}}
\caption{Dark Higgs production in scalar \a~and vector \b~meson decay through dark Higgstrahlung.
\label{fig:DHprod}}
\end{figure}

The produced meson has a tiny chance of decaying into dark sector particles. In this
decay, dark Higgs boson can be produced from an excited
dark photon through a ``dark'' Higgstrahlung mechanism. The processes for the scalar meson decay are
$$\pi^0,\eta \rightarrow \gamma V^*, V^* \rightarrow S
V,$$ and for the vector meson case $$\rho,\omega \rightarrow  V^*, V^* \rightarrow S V.$$
The corresponding Feynman diagrams are shown in
Fig.~\ref{fig:DHprod}.

Focusing on the first process, we can write the branching ratio for a neutral pion as 
\begin{align}
 \text{BR}_{\piz \rightarrow \gamma S V} = \frac{1}{2 m_{\piz} \Gamma_\piz} \  \int \frac{ds}{2\pi} d\Pi_{\piz\rightarrow \gamma V^* } d\Pi_{V^*\rightarrow  V S } |\scr{M}|^2 \ ,
\end{align}
where $d\Pi_{\piz\rightarrow \gamma V^* }$ and  $d\Pi_{V^*\rightarrow V S }$ represent
the usual two-body decay phase space, $|\scr{M}|^2$ is the squared, averaged amplitude, $s$
is the squared momentum of the excited dark photon and is integrated between
$(\mzp+\ms)^2$ and $m^2_{\piz}$. The relevant quantity for our Monte-Carlo simulation is
the differential decay rate $\frac{d \text{BR}_{\piz \rightarrow \gamma S V}}{ds d
\theta} $, where $\theta$ is the angle between the dark Higgs boson and the excited dark photon
in the rest frame of the latter.  We find (see Appendix~\ref{sec:app} for details)
\begin{align}
  \frac{d^2 \text{BR}_{\pi^0 \rightarrow \gamma S V}}{ds d \theta} = \text{BR}_{\pi^0 \rightarrow \gamma \gamma} \times \frac{\eps^2 \azp q_S^2}{8\pi}  \ s \left(  1 - \frac{s}{m^2_{\piz}}\right)^6  \times \frac{ \sqrt{\lambda}~(8 \mzp^2/s + \lambda \sin^2 \theta )}{(s-\mzp^2)^2 + \mzp^2 \Gamma^2_V}\sin \theta \ ,
\end{align}
where $q_S$ is the dark Higgs boson $U(1)_D$ charge, $\Gamma_V$ is the width of the dark
photon (which can be neglected in practice) and $\lambda$ is given by
\begin{align*}
 \lambda ~\equiv~ \left(1 - \frac{(\mzp+\ms)^2}{s}\right)\left(1 - \frac{(\mzp-\ms)^2}{s}\right) \ .
\end{align*}
The case of the $\eta$ meson is completely similar, with the replacement $m_{\pi^0} \rightarrow
m_{\eta}$ and $\text{BR}_{\piz \rightarrow \gamma \gamma} \rightarrow \text{BR}_{\eta \rightarrow \gamma \gamma} = 0.394$. We have also checked agreement with
the integrated standard results of\cite{Batell:2009di}.

The second process, corresponding to vector meson decays, is a simpler two-body decay. The branching ratio is given by
\begin{align}
\text{BR}_{\rho \rightarrow S V} = \text{BR}_{\rho \rightarrow e^+ e^-} \frac{ \eps^2 \azp q_S^2}{\alpha_{em}} \  m^4_{\rho} \frac{ \sqrt{\lambda'}~(12 \mzp^2/m^2_{\rho} + \lambda'  )}{(m^2_{\rho}-\mzp^2)^2 + \mzp^2 \Gamma^2_V} \ ,
\end{align}
where
\begin{align*}
 \lambda'  ~\equiv~ \left(1 - \frac{(\mzp+\ms)^2}{m^2_{\rho}}\right)\left(1 - \frac{(\mzp-\ms)^2}{m^2_{\rho}}\right) \ ,
\end{align*}
and similarly for $\omega$ mesons.

While the processes described above are typically suppressed compared to the on-shell
production of dark matter particles from dark photon decay, the dark Higgs boson on the
other hand is easier to detect as one can search directly for its decay products.
Note that due to the
absence of gauge vertices between two dark Higgs bosons and the dark photon, the only
scattering process available is through dark Higgs boson mixing with the SM Higgs boson and is therefore negligible here.

\subsubsection{Dark Higgs boson decay and detection}
\label{subsubsec:detection}

As discussed before, 
when the decay of a dark Higgs boson into two dark photons or dark matter particles is kinematically
forbidden, it is long-lived and can only decay to an $e^+e^-$ pair through
the loop-diagram process shown in Fig.~\ref{fig:tauA}. This is in principle a very
distinctive signature compared to dark matter scattering. In practice
however, most of the existing experimental bounds are derived from neutrino-electron
scattering signal, which consist of only one charged track. In detail, for each of the
considered experiments, we have:
\begin{itemize}
 \item LSND:  We choose to use the search\cite{Athanassopoulos:1997er} for electron
	 neutrino $\nu_e$ via the inclusive charged-current reaction $\nu_e + C
	 \rightarrow e^- + X$.\footnote{Note that this is not the
	 search\cite{Aguilar:2001ty} which focused on the lower energy region $18
	 \text{ MeV} < E_{e^+} + E_{e^-} < 50 \text{ MeV}$ used, e.g. in\cite{Izaguirre:2017bqb}. 
	 The cut on the electron energy made it
	 unsuitable for our setup.}  Following\cite{Essig:2009nc}, we will consider that
	 the outgoing $\epem$ pair is interpreted as a single electron event
	 satisfying the energy cut, $60 \text{ MeV} < E_{e^+} + E_{e^-} < 200 \text{ MeV}$
	 and use the electron detection efficiency of around $10\%$. Given the
	 uncertainties presented in\cite{Athanassopoulos:1997er} (see especially Fig $29$
	 and the Tables IV and V), and the fact that the energy distribution of our
	 process would not have been uniform, we will consider that $25$ events should
	 have been observed and draw our contours accordingly. As was already pointed out
	 in\cite{Essig:2009nc} for dark photon searches, a re-analysis of the LSND data
	 focused on pair of $e^+e^-$ events and increasing the energy threshold
	 would significantly improve the limit from this experiment. 
 \item MiniBooNE: We concentrate on the ``off-target'' dataset used
	 in\cite{Aguilar-Arevalo:2017mqx} for dark matter searches, and therefore require
	 the electron and positron tracks to satisfy $\cos \alpha > 0.99$ where $\alpha$
	 is the angle to the beam axis and have energy in $50 \text{ MeV} < E_{e^\pm} <
	 600 \text{ MeV}$. The efficiency for detecting leptons is taken to be $35\%$
	 from\cite{Patterson:2009ki}. Following\cite{Izaguirre:2017bqb}, we will require
	 that both leptons are sufficiently separated so that miniBooNE could resolve both
	 tracks (with a angular resolution of $2^{\circ}$). Since no such search has been
	 yet released, we can only give projections.
 \item SNBD: We will conservatively apply the same lepton detection efficiency and cut  $\cos \alpha > 0.99$  as in the MiniBooNE analysis for
	 this experiment, as this is enough to significantly extend the reach of
	 MiniBooNE.
\end{itemize}

\begin{figure}[t]
	\centering
\includegraphics[width=0.65\textwidth]{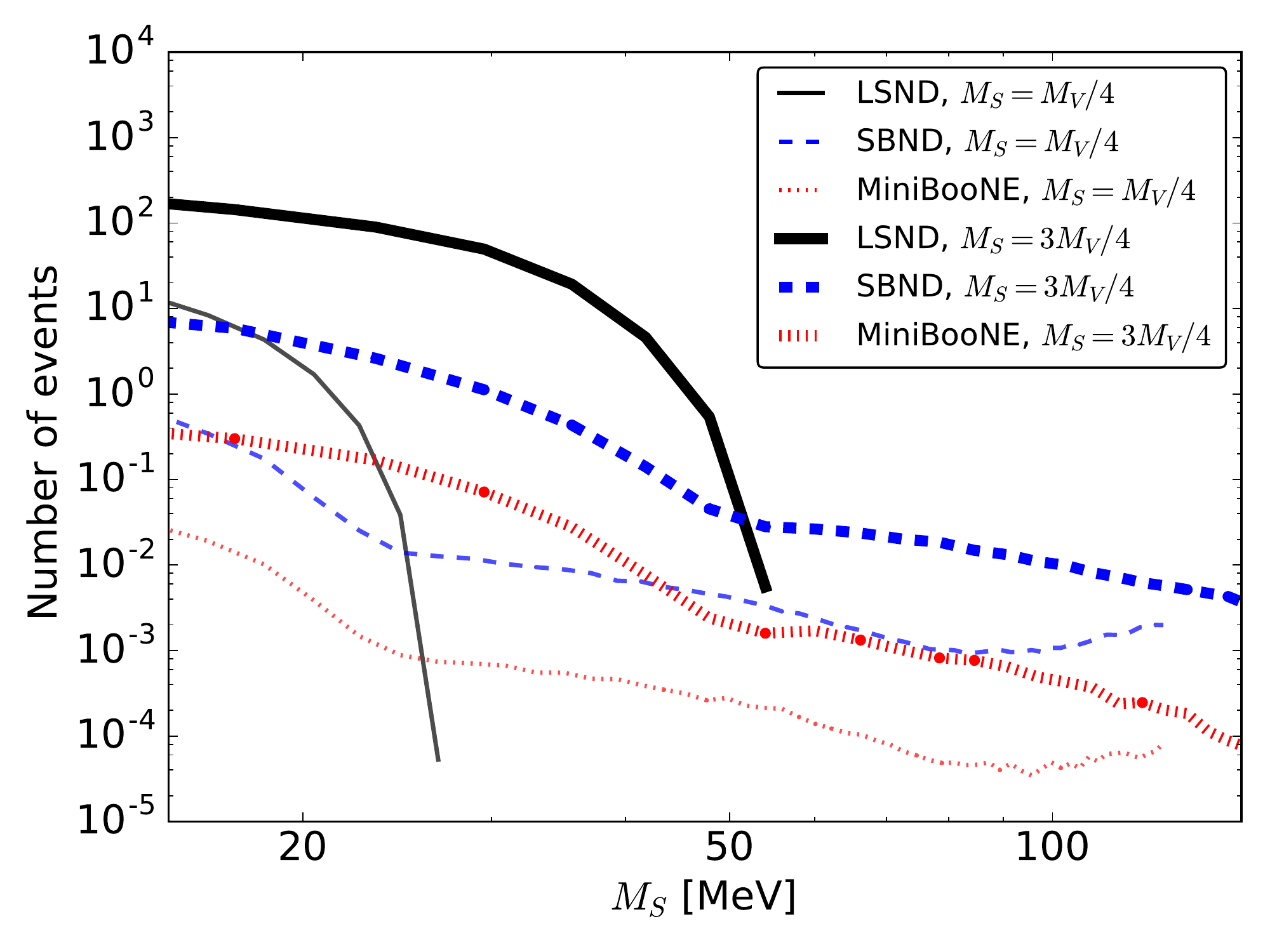}
\caption{Number of events expected at LSND, miniBooNE and SBND experiments as a function of dark Higgs boson mass $\ms$. We show two mass ratios: $\ms =  \mzp/4$ (thin lines) and $\ms = 3/4 \mzp $ (thick lines). We have chosen the couplings to be $\eps = 0.001$ and $\azp = \aem$.
\label{fig:DHNoE}}
\end{figure}

Once the dark Higgs bosons have been produced, they will travel through the shielding before
decaying into the detector. The probability of a
decay event happening within the detector is simply given by
\begin{align}
 \scr{P}_{d} =  \exp \left(-\frac{L_\mathrm{d}}{\gamma v \tau_S} \right) \left[ 1 - \exp \left(-\frac{L_\mathrm{cr}}{\gamma v \tau_S} \right) \right] \ ,
\end{align}
where $\gamma$ is the Lorentz factor, $L_\mathrm{d}$ is the distance between the target and the entry point of the dark
Higgs in the detector and $L_\mathrm{cr}$ is the length of the intersection of the dark
Higgs trajectory with the detector. In the limit where $\gamma v \tau_S \ll
L_\mathrm{d},L_\mathrm{cr}$ the probability reduces to
\begin{align*}
  \scr{P}_{d} \simeq  \frac{L_\mathrm{cr}}{\gamma v \tau_S} \ .
\end{align*}
The number of dark Higgs bosons detected scales as $\eps^6 \azp^2$ so that even
a tiny modification of the kinetic mixing will lead to drastic changes in the detection
signature. We show in Fig.~\ref{fig:DHNoE} the number of events expected in
all three experiments considered here as a function of $\ms$ in our \pdf model. We have chosen
$\eps = 0.001$ and $\azp = \aem$ but the expected number of events for any other
values of these parameters can be recovered from the previously mentioned scaling
relations. In particular, notice
that SBND will improve on the miniBooNE bound by one order of magnitude, provided a
suitable search strategy is implemented. 

\begin{figure}
	\centering
\subfloat[]{\includegraphics[width=0.49\textwidth]{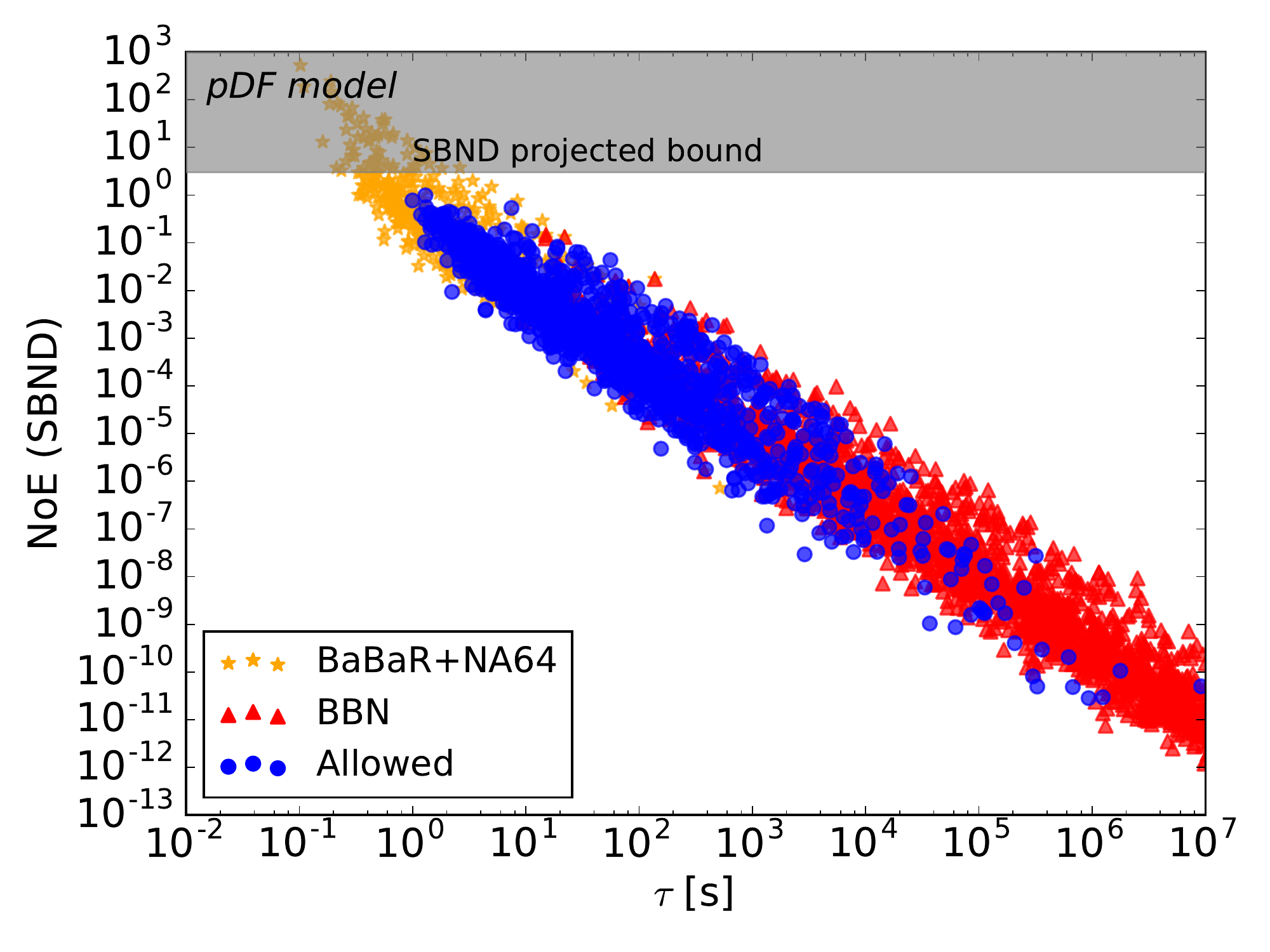}}
\subfloat[]{\includegraphics[width=0.49\textwidth]{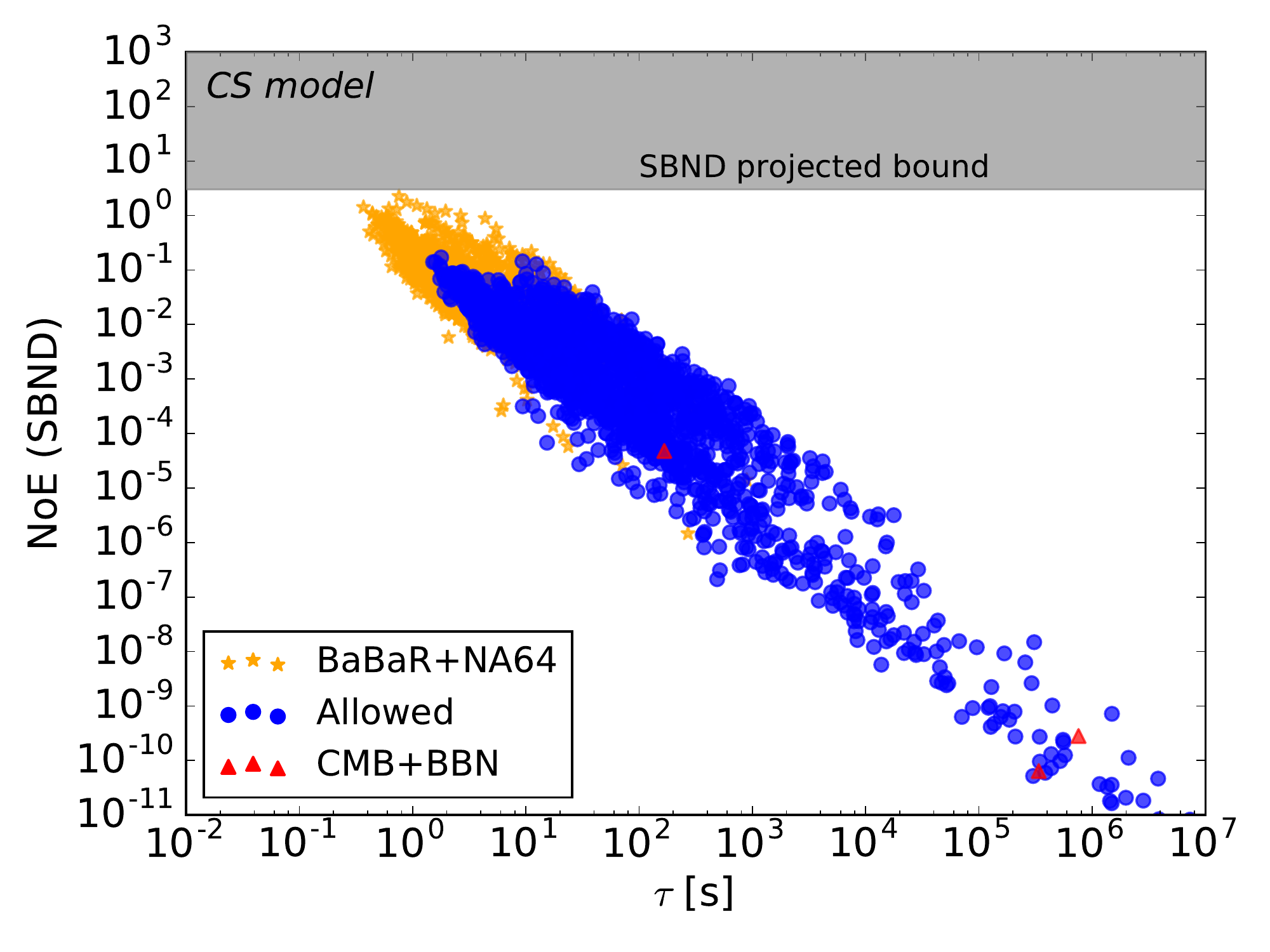}}
\caption{Number of events expected at the SBND experiment for all points satisfying the relic density bound as a function of the dark Higgs boson lifetime $\tau_S$. We show as orange stars all the points excluded by the missing energy searches, and as red triangles points excluded by relevant CMB and BBN observables. We show the reach of SBND assuming no event  observed in the zero background hypothesis for the \pdf~case \a~and the \cs~case \b. The exclusion line is therefore drawn for $95 \%$ CL assuming a Poisson distribution ($3$ events).}
\label{fig:NoE_tau}
\end{figure}

\begin{figure}
	\centering
\subfloat[]{\includegraphics[width=0.49\textwidth]{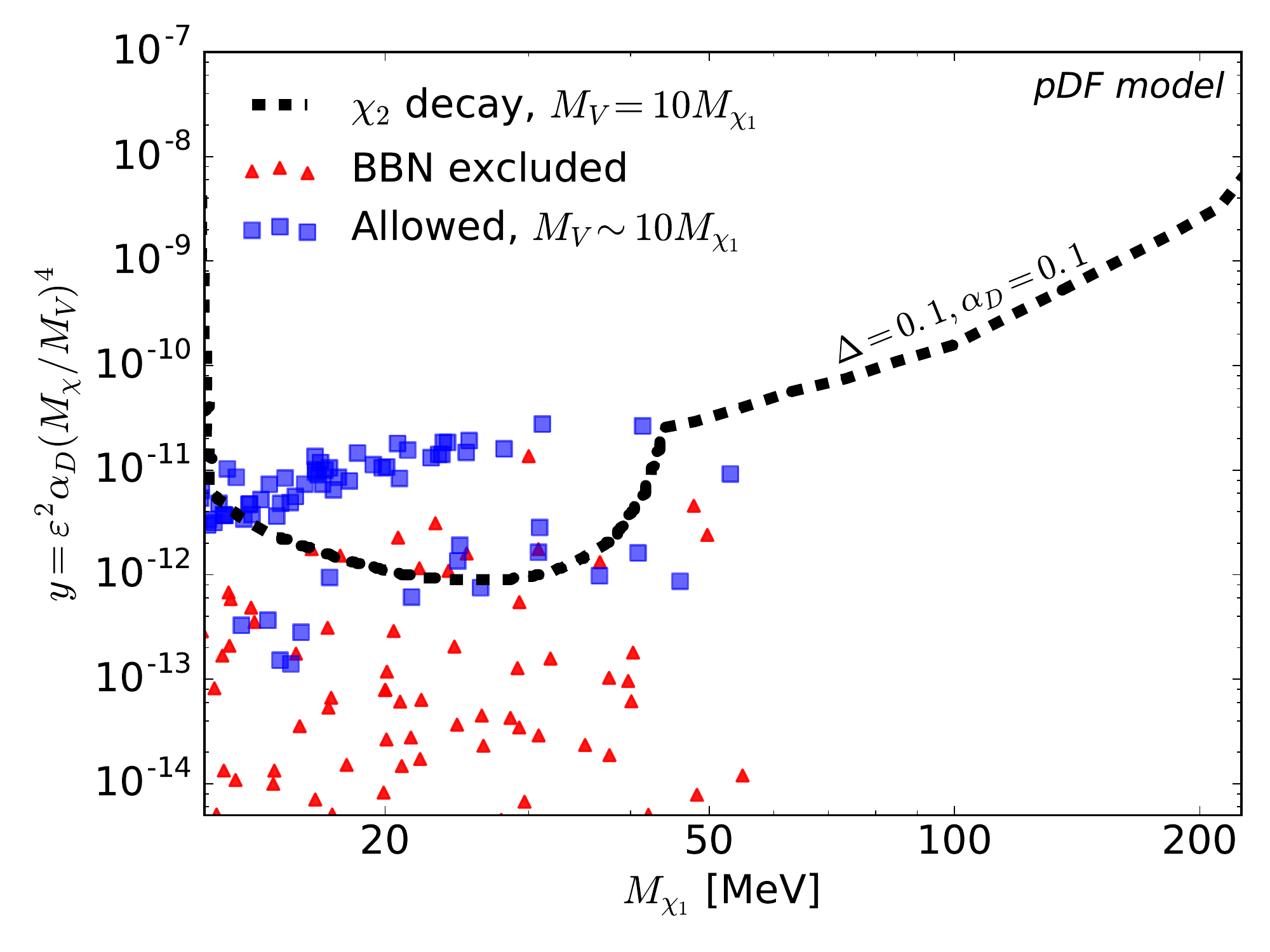}}
\subfloat[]{\includegraphics[width=0.49\textwidth]{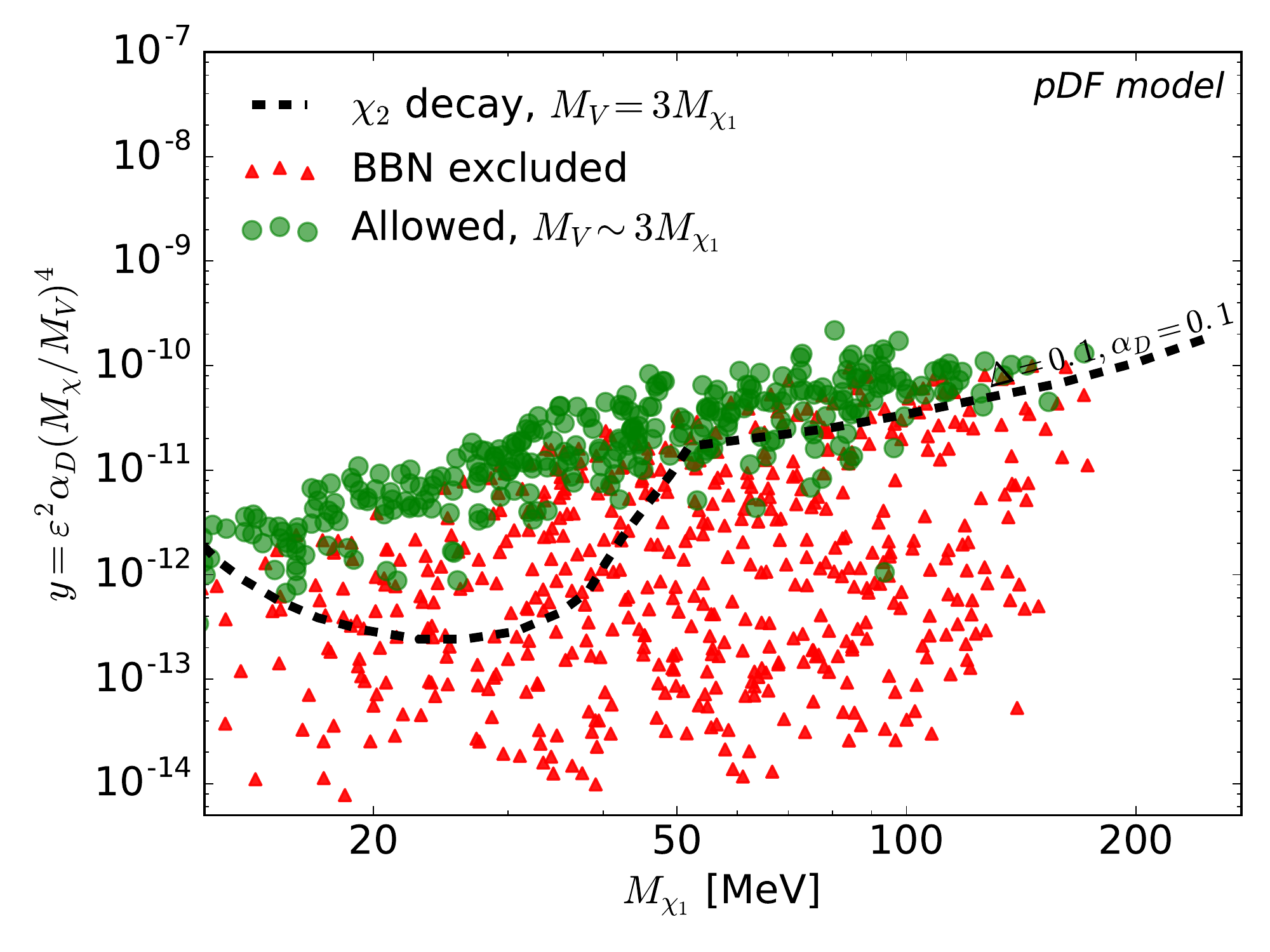}}
\caption{An example of bounds on the parameter $y \equiv  \varepsilon^2 \azp (M_\chi/M_V)^4 $ as a function of the DM mass $\mdm$  for $\Delta \sim 0.1,\azp \sim 0.1$ from\cite{Izaguirre:2017bqb}. We show the points from our scans satisfying all our constraints as well as $\Delta < 0.1,\azp < 0.1$  and  $\mzp \sim 3 M_{\chi_1} $ \a~and $\mzp \sim 10 M_{\chi_1} $ \b. Since the bound is weaker for smaller $\Delta$ and $\azp$, the represented lines are the strongest possible bounds from the analysis of\cite{Izaguirre:2017bqb} for both sets of points.}
\label{fig:yDM}
\end{figure}

Compared with the standard bounds from dark matter searches in this experiment, as in,
e.g.\cite{Aguilar-Arevalo:2017mqx,deNiverville:2016rqh,Izaguirre:2017bqb}, our expected
number of events is even more sensitive to the kinetic mixing parameter $\eps$. In
both of our models, we found that the thermal value target is still out of reach of beam
dump experiment as shown in Fig.~\ref{fig:NoE_tau}, where we have shown the projected
number of events at SBND. The cases of LSND and miniBooNE are similar, with no points
compatible with the relic density constraint leading to more than a few expected events. 
Hence the situation for dark Higgs boson search at proton beam-dump experiments is relatively similar to the one for the dark matter scattering searches in the same detectors, with the thermal value target out of reach of current experiments\cite{deNiverville:2016rqh}. One interesting exception in the \pdf~case was pointed out in\cite{Izaguirre:2017bqb}. When dark matter is produced from dark photon decay, the heaviest mass eigenstate $\chi_2$ can only decay to $\chi_1$ through an off-shell dark photon, in the process $\chi_2 \rightarrow \chi_1 e^+ e^-$, which leads to a long lifetime of order
\begin{align}
\tau_{\chi_2} \sim 3 \cdot 10^{3} \textrm{ m} \times \left( \frac{\aem}{\azp}\right) \left(  \frac{ 0.1}{\Delta}\right)^5 \left(  \frac{ 10^{-3}}{\eps}\right)^2 \left( \frac{75 \textrm{ MeV}}{m_{\chi_1}}\right)^5 \left( \frac{\mzp}{ 200 \textrm{ MeV}}\right)^4 \ ,
\end{align}
where we have introduced the splitting parameter between the two dark matter eigenstate $\Delta = (M_{\chi_2}-M_{\chi_1})/M_{\chi_1}$. While this is not long-lived enough to imply sizable constraints from BBN-related observables, one can search for the $e^+e^-$ pair produced by the decay. The reach is then significantly stronger as we show in Fig.~\ref{fig:yDM}. However, their bounds depends significantly on $\Delta$ and is rapidly not competitive for lower values.   

In\cite{Izaguirre:2017bqb} the thermal value target was almost systematically excluded for dark matter masses in our range of interest, however the fact that the dark Higgs boson opens several new annihilation channels modifies strongly this prediction, as we show in Fig.~\ref{fig:yDM}. Furthermore, the presence of a light dark Higgs boson modifies even more significantly the phenomenology when $M_{\chi_2}-M_{\chi_1} > \ms$. Indeed, the structure of our Lagrangian (namely the
possibility of different Yukawa couplings between the right-handed and left-handed part of
the original dark matter Dirac field) allows for the unsuppressed decay $\chi_2
\rightarrow \chi_1 S $. Hence, in this particular regime, the previous search channel is
no longer open, but should be replaced by a search for dark Higgs boson decay as described
above. This production mechanism should however be orders of magnitude larger than the
Higgstrahlung, as it proceeds completely on-shell, leading to much stronger bounds than
the one from Fig.~\ref{fig:NoE_tau}. Thus, it would be interesting to re-run the search presented in\cite{Izaguirre:2017bqb} (in particular by estimating upcoming bounds from BDX\cite{Battaglieri:2016ggd}) while including the effect of a light dark Higgs boson. We save this analysis for future work.

\subsection{BBN constraints}
\label{subsec:BBN}

Bounds on dark Higgs bosons from BBN can be
surprisingly strong, limiting lifetime to be as small as $0.1\,$s for sub-GeV dark
Higgs when mixing with the SM Higgs boson is considered, as shown in\cite{Fradette:2017sdd}. As we will show in this section, these
constraints will be mitigated in our case due to two factors. First, due to its small mass, the dark Higgs boson
decays only leptonically during BBN, and second, the annihilation mechanisms for our
$\udark$-charged Higgs boson are significantly more effective than the one
in\cite{Fradette:2017sdd}, so that the metastable density of dark Higgs boson after
freeze-out is orders of magnitude smaller.

The decay
products of long lived particles like the dark Higgs boson during the evolution of the Universe can distort the
agreement between the standard BBN predictions and experimental observations of
primordial abundances of light nuclei, in particular $^3$He and D. However, the annihilation of dark Higgs bosons during freeze-out provides a mechanism for depletion
that can in turn ameliorate this potential disagreement.  The energy injections from the decay of
such long lived particles can be at early or late time.  Here, early time refers to the early stages of
BBN when $t\lesssim 10\,$s, wherein decays from a long lived particle could affect the
neutron to proton ratio, $n/p$, or the effective number of neutrino species, $\neff$.
Late time refers to the later stages of BBN when $t\gtrsim 100\,$s which affects the final
primordial abundances of light nuclei.  We shall discuss constraints from both early as
well as late time energy injection from $S$ decays.

\begin{figure}[t!]
	\centering
 
\subfloat[]{\includegraphics[width=0.49\textwidth]{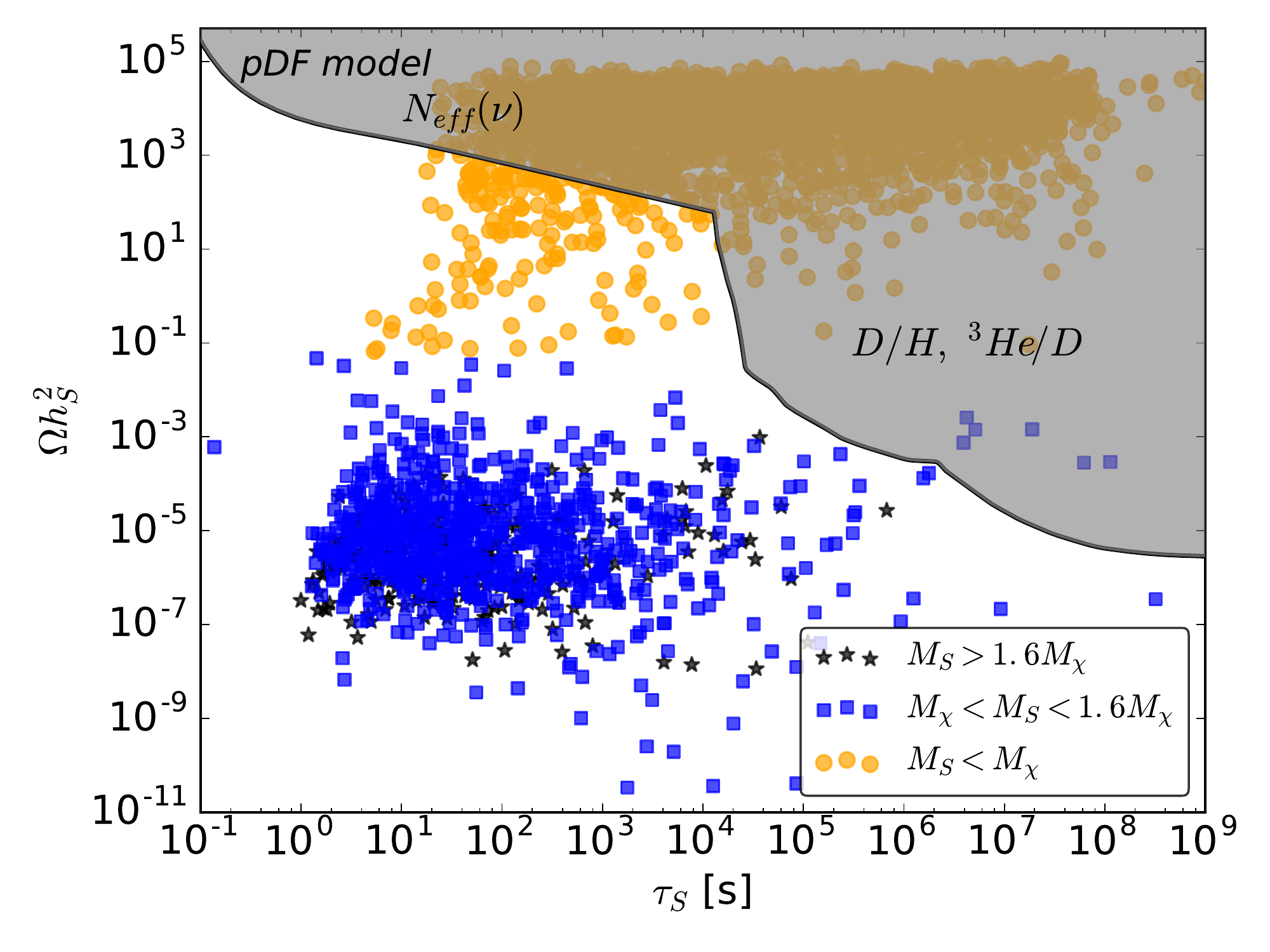}}
\subfloat[]{\includegraphics[width=0.49\textwidth]{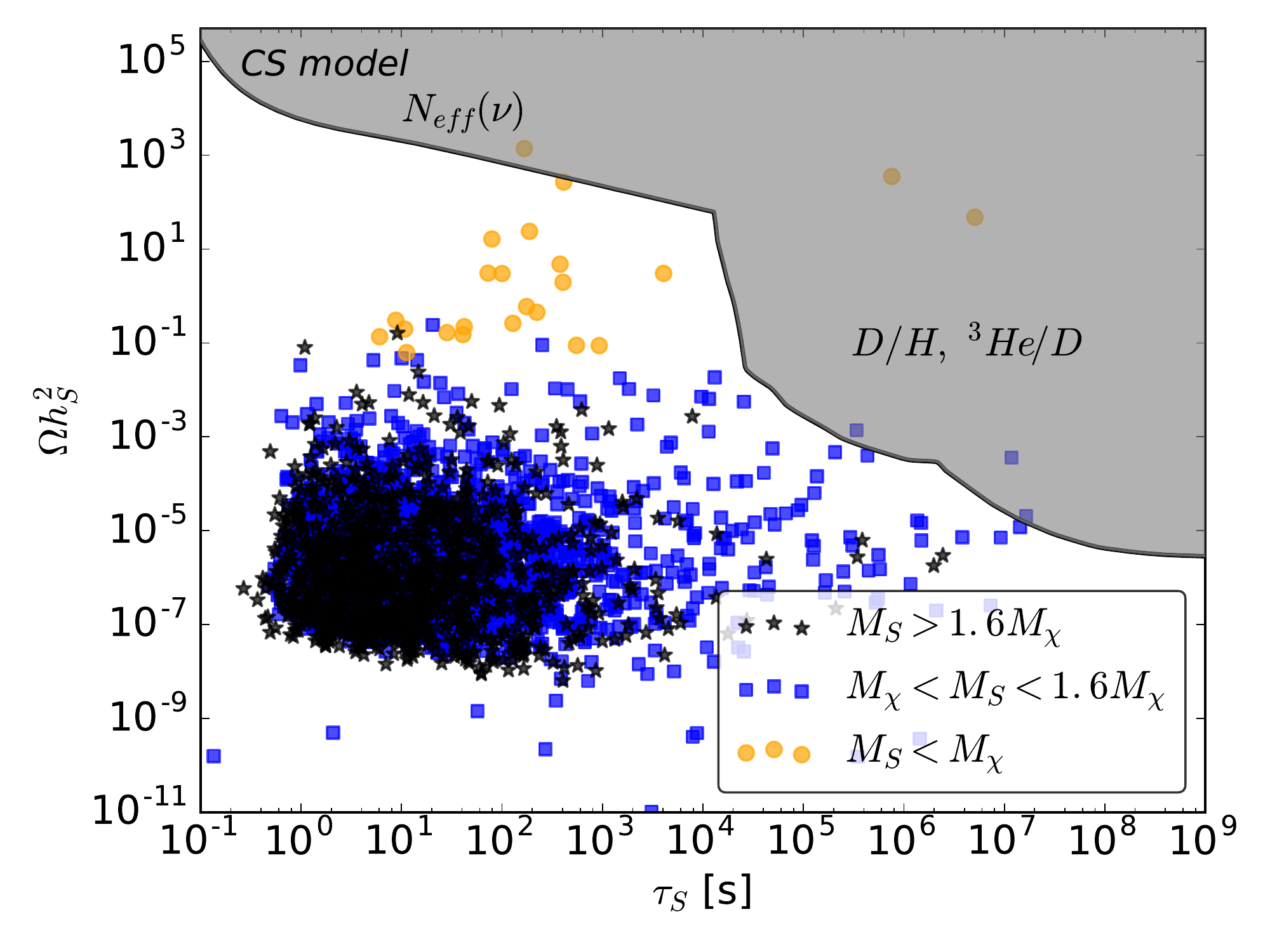}}
\caption{Constraints from light element abundances and from the effective number of neutrinos on the lifetime of the dark Higgs boson $\tau_S$ and on its metastable relic density after freeze-out. In order to allow simple comparison with the dark matter relic density, we show the relic density $\Omega h^2_{S}$ the dark Higgs boson would have had today if it was stable. Points satisfying the dark matter relic density constraint are overlaid for the \pdf~model \a~and for the \cs~case \b. The points have been sorted according to the mass range of the dark Higgs boson. Notice that CMB-related bounds are not included, which explain why points with $\ms < \mdm$ remain in the \cs~case.} 
\label{fig:bbn}
\end{figure}

First we consider constraints from energy injection at early time.  In particular,
hadronic decays of dark Higgs boson, like for example mesons, occurring in the early
universe could significantly alter the $n/p$ ratio.  Similarly, the direct production of
neutrons and protons through quarks and gluons when the dark Higgs boson is sufficiently heavy
can also give rise to stringent constraints on the lifetime of the dark Higgs boson\cite{Cyburt:2002uv,Kawasaki:2004yh,Jedamzik:2006xz,Pospelov:2010hj,Kawasaki:2017bqm}.
However, in our case we restrict the dark Higgs boson mass $M_S$ to be less than the
dimuon threshold.  This also means that there are no hadronic modes available for the dark
Higgs boson decay and the only possible decay mode involves electrons.   As a result we
avoid stringent constraints from hadronic injections and instead we concentrate on the
effect of injection of electrons from $S$ decay.  The effect it can have on BBN can be
constrained using the PLANCK measured value of $\neff$.  The definition of effective
number of neutrino species assumes that the three neutrino species instantaneously
decouple giving a definite neutrino-photon temperature ratio $T_\nu/T_\gamma=(4/11)^{1/3}$,
so that
\begin{equation} 
	\neff=N_\nu\left(\frac{T_\nu}{T_\gamma}\right)^4\left(\frac{4}{11}\right)^{-4/3} \ ,
	\label{neff} 
\end{equation}
where $N_\nu=3$ is the number of neutrino species.  Since the energy injected by the $S$
decays will lead to a reheating of the electron-photon bath with respect to the neutrinos,
this decreases $T_\nu/T_\gamma$.  And as can be seen from Eq.~\eqref{neff}, this leads to a
lowering of $\neff$.  We use the result obtained in ref.\cite{Fradette:2017sdd} where an
approximate analytical approach was adopted to calculate $\neff$ assuming a neutrino
decoupling temperature of $1.4$ MeV.   The $2\sigma$ lower bound from PLANCK\cite{planck2015} requires $\neff>2.71$. We show in
Fig.~\ref{fig:bbn} the exclusion limit on dark Higgs boson
lifetime, $\tau_S$, as a function of $\Omega_S h^2$, the relic density the dark Higgs boson would have had today if it was stable.  We see that most of the parameter space survives as a result of
efficient annihilation channels of dark Higgs boson, particularly when $\ms>\mdm$ and the dark
Higgs boson can annihilate into dark matter thereby decreasing its abundance substantially.
However, when $\ms<\mdm$ this annihilation channel is not so efficient and the metastable
abundance can be quite large and some of the parameter space especially above
$\tau\sim100\,$s is ruled out.

Next we consider the effect of late time energy injections at $t\gtrsim 100\,$s.  Such late energy
injections can potentially destroy light nuclei through dissociation thereby altering their
abundances.  When the long lived particle primarily decays to hadrons the resulting
hadro-dissociation can be very effective in reducing the primordial abundances even at relatively
early times $t\sim 100\,$s.  But once again in the dark Higgs boson scenario considered here the only
viable decay mode is $e^+e^-$ which leads to constraints only from electromagnetic showers.  The
absence of hadronic showers means that we avoid severe constraints from measurement of primordial
abundances.  The constraints from electromagnetic showers arise through photo-dissociation of light
nuclei which become significant at $t\gtrsim 10^4\,$s.  At $t\sim 10^4 - 10^6\,$s, the
photo-dissociation of deuterium, while at $t\gtrsim 10^6$ the over production of D and $^3$He through
the photo-dissociation of $^4$He lead to the most stringent constraints\cite{Kawasaki:2017bqm}. The choice of parameters in
this case as mentioned in the next section, leads to a lifetime in the range of $1-10^5 $ s.  In this
range of dark Higgs boson lifetime the bounds from $\neff$ are the most stringent up to
$\sim 10^4\,$s, and
above $10^4\,$s the bounds from D/H and $^3$He/D are the most stringent as far as BBN is concerned.
In the dark Higgs boson scenario, however, there can be additional annihilation channels which can reduce
the metastable abundance as mentioned in Sec.~3.1. For example, the production of a dark matter pair
from dark Higgs boson annihilation can be significant for $\ms\simeq\mdm$, thereby potentially avoiding
constraints from BBN.  In Fig.~\ref{fig:bbn} we show the exclusion limits from D/H and
$^3$He/D abundances.  We see that most of the parameter space above $10^4\,$s is ruled out
by these bounds, however one could still have substantial annihilation into dark matter which may
allow a few points in the parameter space especially in the $\pdf$ case.

Bounds on the lifetime translate almost directly into a lower bound for the
kinetic mixing parameter from Eq.~\eqref{eq:Slifetime}. When the dark Higgs boson
metastable density is large as no effective annihilation into dark matter is possible,
then we have the rough bound $\eps \gtrsim 10^{-4}$. 
When $\ms \gtrsim \mdm$ the metastable density is
suppressed by the annihilation process $S S \rightarrow \chi \chi$ which dominates over the reverse process, and the bounds are significantly weakened. 
Most points still have $\tau_S \lesssim 10^{4}$ s as can be seen in Fig.~\ref{fig:bbn}. However, this is not a strong bound and some more fine-tuned points can have longer lifetime, of order $10^6$ s. For
such high values of $\tau_S$, the mixing with the SM Higgs boson (which we neglected
following the discussion of Sec.~\ref{sec:lifetime}) should become competitive to mediate the
dark Higgs boson decay. 

\section{Summary and conclusions}
\label{sec:res}

We have argued that in models with a massive, but light, dark vector mediator, the spectrum should
naturally contain a light dark Higgs boson, whose presence can substantially modify the
predictions of the two
models considered in this paper. In the plane $\mdm - \ms$ we can identify four
regions, as shown in Fig.~\ref{fig:regions}, each with very distinct phenomenologies:
\begin{figure}[t]
	\centering
\includegraphics[width=0.8\textwidth]{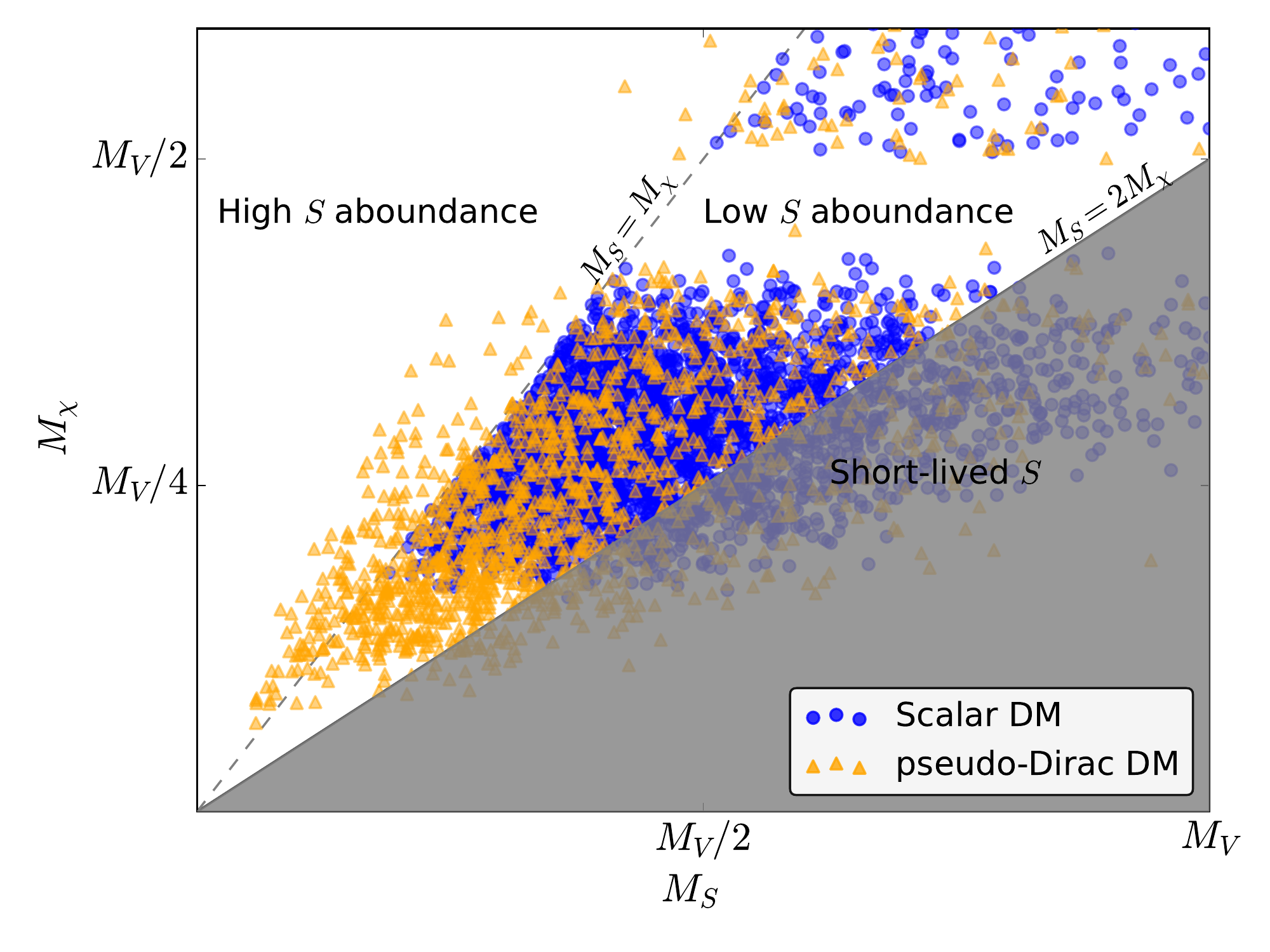}
\caption{The four phenomenologically distinct regions described in the text shown in the dark matter mass $\mdm$ versus the dark Higgs boson mass $\ms$ plane with data points from the \cs~case (blue) and the \pdf~case (orange) which satisfy all the constraints considered in this analysis. The grey-shaded regions mark the two regimes where the phenomenology does not significantly differ from previous studies.
\label{fig:regions}}
\end{figure}
\begin{itemize}
 \item The \textit{secluded} regime ($M_\chi \gtrsim \mzp $) in which dark matter annihilation into $VV$ becomes relevant. This tends to wash out the relic density for the value of the dark gauge coupling considered, but is furthermore heavily constrained by CMB bounds as this is an $s$-wave process. We observed almost no points from our scans in this region.
 \item The \textit{short-lived dark Higgs boson} regime corresponding to relatively heavy
	 dark Higgs boson. This is the case considered in most of the previous literature,
	 most notably recently in\cite{Izaguirre:2017bqb} for the \pdf~model. Dark Higgs
	 bosons tend to decay instantaneously into a dark matter pair, leaving little new
	 imprint, both in beam-dump experiments and in cosmological observables.
\item The \textit{long-lived dark Higgs boson} regime in which the dark Higgs boson is light
	enough so that it cannot decay into dark photon or dark matter particles. Its
	decay products can then be observed in beam-dump experiments, even though the
	corresponding bounds are often weaker than the missing energy searches by BaBar
	and NA64. Depending on whether or not one has $\ms \lesssim \mdm $, this regime
	divides into two sub-regions:
\begin{itemize}
 \item The low abundance region, $\mdm < \ms < 2 \mdm$, where the process $S S \rightarrow \chi \chi  $ is
	 effective. The metastable density of dark Higgs bosons after freeze-out is
	 therefore suppressed, so that the bounds from BBN are weakened.  
 \item The high abundance region, $\ms < \mdm$, where there is no effective annihilation process for
	 the dark Higgs boson. The consequent high metastable density of dark Higgs bosons
	 translates into relatively strong bounds from BBN-related observables.
	 Furthermore, the dark matter annihilation channel $\chi \chi \rightarrow S S$  is
	 kinematically open. Being an $s$-wave process for the model \cs, this region is
	 ruled out by CMB constraints, as can be seen in
	 Fig.~\ref{fig:regions}. 
\end{itemize}
\end{itemize}
\begin{table}[t]
\begin{center}
\begin{tabular}{|m{4.cm}|m{2.25cm}|m{2.25cm}|m{2.25cm}|m{2.5cm}|}
\hline
\rule{0pt}{1.25em}
   & \multicolumn{2}{c|}{\textbf{\pdf~model}} & \multicolumn{2}{c|}{\textbf{\cs~model}} \\[0.15em]
   \hline
   \rule{0pt}{1.5em}\textbf{Parameter} &  \textbf{Low  \abunS } & \textbf{High \abunS } &  \textbf{Low \abunS } & \textbf{Short-lived S}\\[0.15em]
\hline
\hline
\rule{0pt}{1.em}
\centering $\lam_S$ & $0.14$ & $ 1.8 \cdot 10^{-3} $ & $1.35 \cdot 10^{-2}$ & $0.09$ \\
\hline
\rule{0pt}{1.em}
\centering $\gzp$ & $0.86$ & $0.23$ & $ 0.46 $& $ 0.49 $ \\
\hline
\rule{0pt}{1.em}
\centering$\mzp$ & $223$ & $73$ & $40$ & $154$\\
\hline
\rule{0pt}{1.em}
\centering$\eps$  & $8.4 \cdot 10^{-4}$ & $6.2 \cdot 10^{-4}$ &  $2.8 \cdot 10^{-4}$ & $4 \cdot 10^{-5}$\\
\hline
\hline
\rule{0pt}{1.em}
\centering$m_\chi$  & $47$ & $15.1$ &  $12.6$ & $58.6$\\
\hline
\rule{0pt}{1.em}
\centering$\lambda_{S\chi}$  & \textendash & \textendash &  $5.2 \cdot 10^{-3}$& $0.016$\\
\hline
\rule{0pt}{1.em}
\centering$\lambda_{\chi}$  & \textendash & \textendash &  $1.8 \cdot 10^{-3}$ & $4.4 \cdot 10^{-3}$\\
\hline
\rule{0pt}{1.em}
\centering$y_{S L}$  &  $0.013$  & $1.63 \cdot 10^{-3}$ &  \textendash & \textendash \\
\hline
\rule{0pt}{1.em}
\centering$y_{S R}$  &   $6.2 \cdot 10^{-3} $  & $2.1 \cdot 10^{-3} $ &  \textendash & \textendash \\
\hline
\hline
\rule{0pt}{1.em}
\centering$\ms$ &  $62.7$  & $9.3$ &   $14.2$ &  $133$ \\
\hline
\rule{0pt}{1.em}
\centering$\mdm$ &   $48.0$ & $15.0$ &  $13.3$ &  $64.8$ \\
\hline
\rule{0pt}{1.em}
\centering \abunS &  $ 3 \cdot 10^{-6} $  & $267$ &  $0.7 \cdot 10^{-5}$  &  \textendash \\
\hline
\rule{0pt}{1.em}Dominant channels&    & &  & \\
$\rightarrow$ relic density, $\avsigv_{\rm an}$&  $e^+e^-$  & $e^+e^-,S S$  & $SS$  &  $e^+e^-$\\
\hline
\rule{0pt}{1.em}$S$ lifetime (s) &  $2.7$  & $101$ &  $436$  & \textendash \\
\hline
\rule{0pt}{1.em}NoE (LSND) &  \textendash   & $0.04$ &  $0.07$  & \textendash \\
NoE (miniBooNE) &  $ 1.1 \cdot 10^{-3}$  & $6.8 \cdot 10^{-5}$ &   $0.14 \cdot 10^{-4}$     & \textendash \\
NoE (SBND) &   $ 0.094 $  & $4.8 \cdot 10^{-3}$&   $3.2 \cdot 10^{-4}$  &\textendash \\
\hline

\end{tabular}
\caption{Benchmark points for the models analyzed in this work.
Mass-related quantities are given in MeV and $\mev^2$, cross-section times velocity are in cm$^3$/s.}
\label{tab:BP}
\end{center}
\end{table}%

\vspace{1cm}

In Table~\ref{tab:BP}, we give benchmark points for these regions satisfying all the constraints considered in this article.

\begin{figure}[t]
	\centering 
	\subfloat[]{	\includegraphics[width=0.49\textwidth]{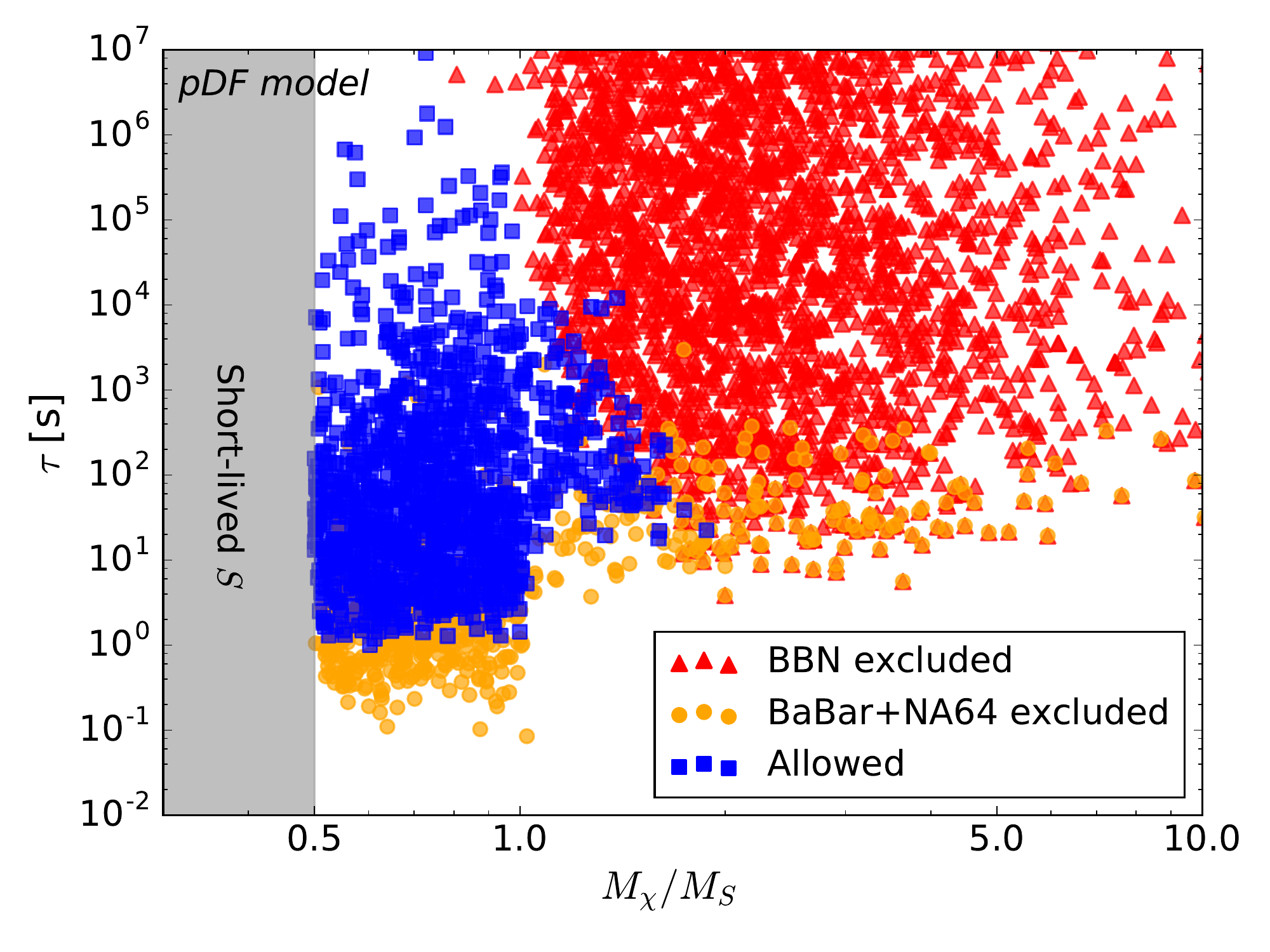} }
	\subfloat[]{\includegraphics[width=0.49\textwidth]{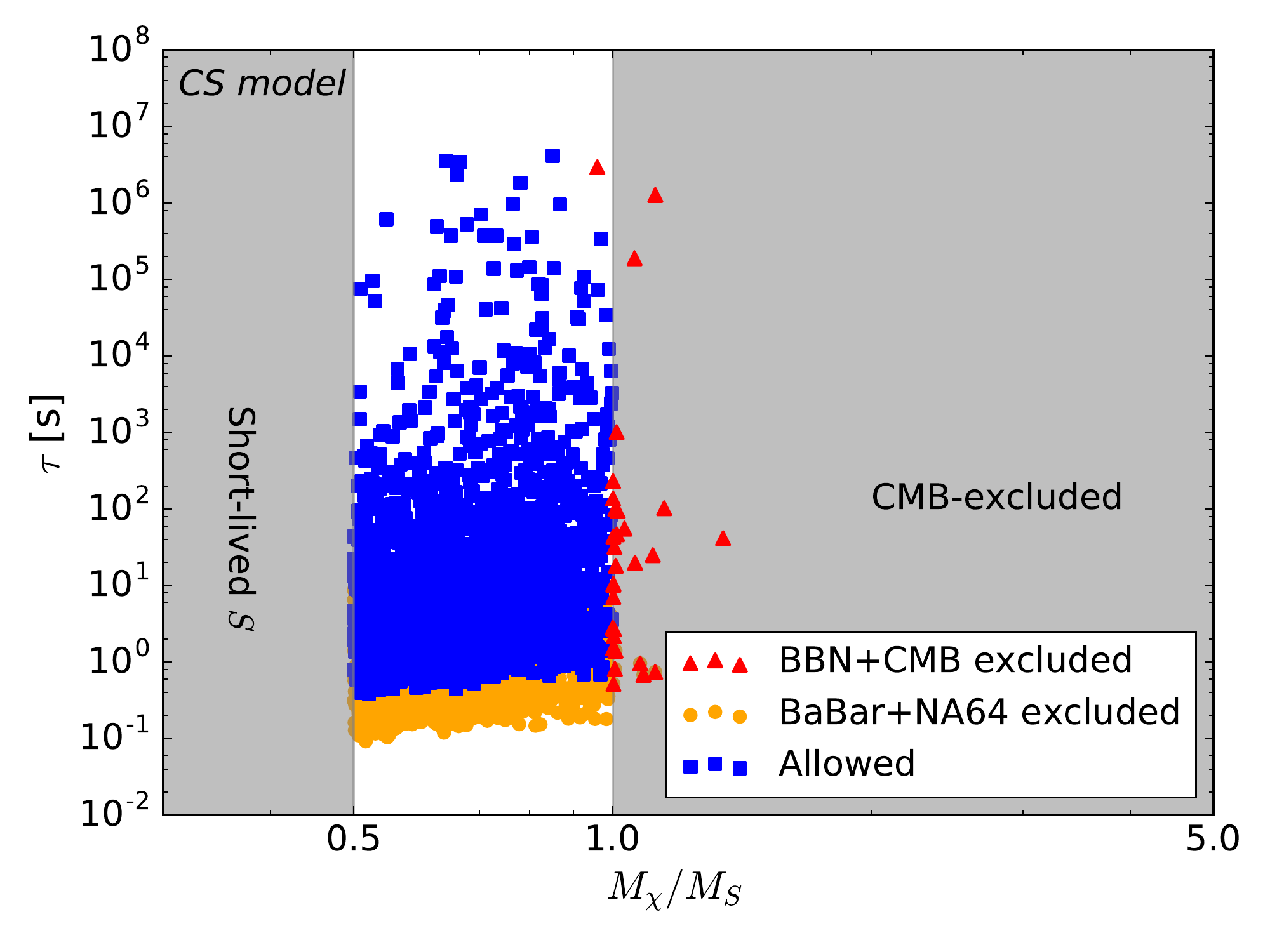}}
	\caption{Summary of the various relevant bounds considered in this analysis, with the points satisfying the relic density constraint in the $\ms / \tau_S$ plane for the
		\pdf~model \a~and  the \cs~case \b~overlaid. We have restricted the points to the region of long-lived dark Higgs boson where the phenomenology is distinctively different from the previous studies of these models.} \label{fig:mstau}
	\end{figure}

In this paper, we have focused on the long-lived dark Higgs boson regime, as the secluded and short
lived dark Higgs boson scenarios had already been covered extensively. We found that while the dark
Higgs boson can in principle be produced and detected in proton beam-dump experiments, the thermal value target is out of reach of the experiments considered here. 
This conclusion should however be mitigated by several
comments. First, as has been already advocated by many previous papers, it would be very interesting
to make a re-analysis of the LSND data, possibly raising the energy threshold for the detected
electrons and looking eventually for $e^+e^-$ pair directly as this will significantly increase the reach of this experiment. Second, our conclusion regarding the
reach of beam-dump experiments only applies to low-energy beam experiments, where the dominant
production mechanism is meson decay. For more energetic beam experiments, or for electron beam dumps, different production channels for the dark Higgs boson, like direct production, should
be considered.  Finally, we expect that dark Higgs boson decay in proton beam-dump experiments could  set stronger bounds  than
those of missing energy searches in the case of the \pdf~model when the process $\chi_2 \rightarrow
\chi_1 S$ is available.

On the other hand, the cosmology of the two models considered
is significantly modified, with additional annihilation channels leading to
various constraints. The
bounds from the CMB arising from the fact that some of the new dark matter
annihilation channels were unsuppressed at recombination time have been presented, excluding
in particular completely the region $\mdm > \ms$ in the \cs~case. Furthermore,
BBN-related observables which arise as a consequence of the long lifetime of the dark Higgs
boson were found to be relevant, but weaker than could have been expected from previous works. We have summarized the main
constraints on both the \cs~ and \pdf models in Fig.~\ref{fig:mstau}.
	
Regarding earth-based experiments, the most promising discovery channels for these models
seem to be the missing-energy searches as they exclude already large portion of the
parameter space. In the case of the \pdf~model, direct detection in beam-dump experiment
of the decay of the heaviest dark matter field as advocated in\cite{Izaguirre:2017bqb} is
a promising strategy which can be further combined with the search for dark Higgs
boson from the $\chi_2 \rightarrow \chi_1 S$ channel when it is kinematically accessible. It
would be interesting to study other types of  cosmological probes for an
extremely long lived dark Higgs boson, as for example possible supernovae-related
constraints or possible signatures from dark Higgs boson (or dark photon) production in DM annihilation in the
sun as was already studied in\cite{Schuster:2009au,Smolinsky:2017fvb} for heavier dark matter candidates.

In the long run, the experimental prospects for both our models are bright. Almost all of the parameter space which meets the thermal value target will be independently probed by the next generation of projected electron beam-dump experiments (for instance LDMX), by direct detection experiments such as XENON1T for the \cs~model, and by indirect detection experiments searching for current dark matter annihilation in the MeV mass range.

\bigskip
\noindent \textbf{Acknowledgments}
\medskip

\noindent The authors warmly thank Felix Kahlhoefer for pointing out an error in the estimation of the dark matter annihilation rates in the pre-print version of this paper. LD and SR would like to thank P. deNiverville for helpful discussions regarding his code BdNMC. LR is supported by the Lancaster-Manchester-Sheffield Consortium for Fundamental Physics under STFC Grant No.\ ST/L000520/1. LD, LR and SR are supported in part by the National Science Council (NCN) research grant No.~2015-18-A-ST2-00748. The use of the CIS computer cluster at the National Centre for Nuclear Research in Warsaw is gratefully acknowledged.

\newpage
\appendix



\section{Differential production rate of dark Higgs boson from meson decay}
\label{sec:app}
 We present in this appendix more details about the differential cross-section corresponding to the three-body scalar meson decays into a photon, a dark photon and a dark Higgs boson (see Fig.~\ref{fig:DHprod}).
 
We associate the four-momentum $p$ to the outgoing photon, $q$ to the excited intermediary dark photon and $k$ the outgoing dark photon. The coupling $c_S$ between a dark Higgs boson and two dark photons is given with our parameter by $$ c_s = \gzp^2 q_S^2 v_S  \ ,$$ where $q_S$ is the dark Higgs boson charge. We use the usual notation $s = q^\mu q_\mu$ and denote the photon (dark photon) polarization four-vector by $\mathbf{e}^\mu$ ($\mathbf{\tilde{e}}^\mu$). Following\cite{Kahn:2014sra} we can then write the amplitude for this process from the one giving the decay of meson into two photons  mediated by the chiral anomaly as
\begin{align}
 \scr{A}_{M \rightarrow \gamma S V} = \frac{\eps \alpha_{em}}{\pi f_\pi}  ( 2  c_S ) \eps^{\mu \nu \alpha \beta} p_\alpha q_{\beta} \mathbf{e}_{\mu} \mathbf{\tilde{e}}^*_{\lambda} \frac{\delta_\nu^{\phantom{\nu}\lambda} - q_\nu q^\lambda / \mzp^2}{(s-\mzp^2) + i \mzp \Gamma_V} \ ,
\end{align}
where we have used the factor $f_\pi $ defined from the decay width of the meson into a pair of photons as
\begin{align}
\Gamma_{M \rightarrow \gamma \gamma} \equiv \frac{1}{f_\pi^2} \frac{\alpha_{em}^2 m_M^3}{(4 \pi)^3} \ .
\end{align}
Using the following useful kinematic relations:
\begin{align*}
 p^2 &= 0 \ , &&& q^2 &= s  \ , &&& k^2 = \mzp^2 \ ,\\
  k\cdot q &= \frac{s + \mzp^2-\ms^2}{2} \ , &&&   p \cdot q &= \frac{m_M^2- s}{2}  \ , &&&  
\end{align*}
we can then square the amplitude and sum over the outgoing polarization states. We  obtain
\begin{align}
\label{eq:amp}
 \langle \left|\scr{A}_{M \rightarrow \gamma S V}\right|^2 \rangle~=~ & \frac{\eps^2 \alpha_{em}^2 c_S^2}{4 \pi^2 f_\pi^2} \frac{1}{(s-\mzp^2)^2 +  \mzp^2 \Gamma_V^2}  \\
 & \quad \left[ \frac{(m_M^2-s)^2}{4}  - \frac{p \cdot k}{\mzp^2} \left( s  \left(p \cdot k \right) - (m_M^2-s)\frac{s + \mzp^2-\ms^2}{2} \right)\right] \ . \nn
\end{align}
Introducing the angle $\theta$ between the outgoing dark Higgs boson and the excited dark photon direction, chosen in the rest frame of the excited dark photon, we can expand $(p \cdot k)$ as
\begin{align*}
 p \cdot k = \frac{1}{4 s } (m_M^2-s) \left[  \left(s+\mzp^2-\ms^2\right) + \sqrt{\lambda \left(s,\mzp^2,\ms^2\right)} \cos \theta \right] \ ,
\end{align*}
where we have used the usual definition for the kinematic triangle function $\lambda$:
\begin{align*}
 \lambda (a,b,c) = a^2+b^2+c^2 -2ab-2ac-2bc \ . 
\end{align*}
In order to simulate the decay chains in our Monte-Carlo simulation, we need the differential branching ratio $\displaystyle \frac{d^2 \text{BR}_{M \rightarrow \gamma S V}}{ds d \theta}$. Writing the two-body phase space differential element $d \Pi^2$, we can use the recursion relations to decompose the three-body phase space into two-body ones combined with an extra integral over the excited dark photon squared momentum $s$, leading to
\begin{align}
 \text{BR}_{M \rightarrow \gamma S V} = \frac{1}{2 m_{M} \Gamma_M} \  \int \frac{ds}{2\pi} d\Pi_{M\rightarrow \gamma V^* } d\Pi_{V^*\rightarrow  V S }\langle \left|\scr{A}_{M \rightarrow \gamma S V}\right|^2 \rangle \ ,
\end{align}
with the integration on $s$ running between
$(\mzp+\ms)^2$ and $m^2_{M}$. Integrating directly on $d\Pi_{M\rightarrow \gamma V^* } $ and on every angle but $\theta$, we have
\begin{align}
\label{eq:kin}
 \int  d\Pi_{M\rightarrow \gamma V^* } d\Pi_{V^*\rightarrow  V S } ~\longrightarrow~  \int  (d\theta \sin \theta) \frac{1}{128 \pi^2} (1-\frac{s}{m_M^2}) \frac{\sqrt{\lambda(s,\mzp^2,\ms^2)}}{s} \ .
\end{align}
Finally, using the definition of  $\mzp \equiv v_S \gzp q_S$, we can combined Eq.~\eqref{eq:amp} and Eq.~\eqref{eq:kin} to get our result
\begin{align}
 \frac{d^2 \text{BR}_{M \rightarrow \gamma S V}}{ds d \theta} = \text{BR}_{M \rightarrow \gamma \gamma} \times \frac{\eps^2 \azp q_S^2}{8\pi}  \ s \left(  1 - \frac{s}{m^2_{\piz}}\right)^6  \times \frac{ \sqrt{\lambda}~(8 \mzp^2/s + \lambda \sin^2 \theta )}{(s-\mzp^2)^2 + \mzp^2 \Gamma^2_V}\sin \theta \ ,
\end{align}
where we used the short-hand notation $\lambda \equiv \lambda \left(1,\mzp^2/s,\ms^2/s \right)$.

\newpage

\bibliographystyle{utphys}
\bibliography{bdump_journal}

\providecommand{\href}[2]{#2}\begingroup\raggedright\begin{thebibliography}{10}

\bibitem{Roszkowski:2017nbc}
L.~Roszkowski, E.~M. Sessolo, and S.~Trojanowski, ``{WIMP dark matter
  candidates and searches - current issues and future prospects},''
\href{http://arxiv.org/abs/1707.06277}{{\ttfamily arXiv:1707.06277 [hep-ph]}}.

\bibitem{Plehn:2017fdg}
T.~Plehn, ``{Yet Another Introduction to Dark Matter},''
\href{http://arxiv.org/abs/1705.01987}{{\ttfamily arXiv:1705.01987 [hep-ph]}}.

\bibitem{Battaglieri:2017aum}
M.~Battaglieri {\em et~al.}, ``{US Cosmic Visions: New Ideas in Dark Matter
  2017: Community Report},'' {\em FERMILAB-CONF-17-282-AE-PPD-T} (2017) ,
\href{http://arxiv.org/abs/1707.04591}{{\ttfamily arXiv:1707.04591 [hep-ph]}}.

\bibitem{Alexander:2016aln}
J.~Alexander {\em et~al.}, ``{Dark Sectors 2016 Workshop: Community Report},''
  {\em FERMILAB-CONF-16-421} (2016) ,
\href{http://arxiv.org/abs/1608.08632}{{\ttfamily arXiv:1608.08632 [hep-ph]}}.

\bibitem{Chu:2016pew}
X.~Chu, C.~Garcia-Cely, and T.~Hambye, ``{Can the relic density of
  self-interacting dark matter be due to annihilations into Standard Model
  particles?},'' \href{http://dx.doi.org/10.1007/JHEP11(2016)048}{{\em JHEP}
  {\bfseries 11} (2016) 048},
\href{http://arxiv.org/abs/1609.00399}{{\ttfamily arXiv:1609.00399 [hep-ph]}}.

\bibitem{Correia:2016xcs}
F.~C. Correia and S.~Fajfer, ``{Restrained dark $U(1)_d$ at low energies},''
  \href{http://dx.doi.org/10.1103/PhysRevD.94.115023}{{\em Phys. Rev.}
  {\bfseries D94} no.~11, (2016) 115023},
\href{http://arxiv.org/abs/1609.00860}{{\ttfamily arXiv:1609.00860 [hep-ph]}}.

\bibitem{Knapen:2017xzo}
S.~Knapen, T.~Lin, and K.~M. Zurek, ``{Light Dark Matter: Models and
  Constraints},''
\href{http://arxiv.org/abs/1709.07882}{{\ttfamily arXiv:1709.07882 [hep-ph]}}.

\bibitem{Kuflik:2017iqs}
E.~Kuflik, M.~Perelstein, N.~R.-L. Lorier, and Y.-D. Tsai, ``{Phenomenology of
  ELDER Dark Matter},'' \href{http://dx.doi.org/10.1007/JHEP08(2017)078}{{\em
  JHEP} {\bfseries 08} (2017) 078},
\href{http://arxiv.org/abs/1706.05381}{{\ttfamily arXiv:1706.05381 [hep-ph]}}.

\bibitem{Feng:2017drg}
J.~L. Feng and J.~Smolinsky, ``{Impact of Resonance on Thermal Targets for
  Invisible Dark Photon Searches},''
\href{http://arxiv.org/abs/1707.03835}{{\ttfamily arXiv:1707.03835 [hep-ph]}}.

\bibitem{Duch:2017khv}
M.~Duch, B.~Grzadkowski, and D.~Huang, ``{Strongly self-interacting vector dark
  matter via freeze-in},''
\href{http://arxiv.org/abs/1710.00320}{{\ttfamily arXiv:1710.00320 [hep-ph]}}.

\bibitem{Feng:2017vli}
J.~L. Feng, I.~Galon, F.~Kling, and S.~Trojanowski, ``{Dark Higgs Bosons at
  FASER},''
\href{http://arxiv.org/abs/1710.09387}{{\ttfamily arXiv:1710.09387 [hep-ph]}}.

\bibitem{Wojtsekhowski:2017ijn}
B.~Wojtsekhowski {\em et~al.}, ``{Searching for a dark photon: Project of the
  experiment at VEPP-3},''
\href{http://arxiv.org/abs/1708.07901}{{\ttfamily arXiv:1708.07901 [hep-ex]}}.

\bibitem{Feng:2017uoz}
J.~Feng, I.~Galon, F.~Kling, and S.~Trojanowski, ``{FASER: ForwArd Search
  ExpeRiment at the LHC},''
\href{http://arxiv.org/abs/1708.09389}{{\ttfamily arXiv:1708.09389 [hep-ph]}}.

\bibitem{Tulin:2017ara}
S.~Tulin and H.-B. Yu, ``{Dark Matter Self-interactions and Small Scale
  Structure},''
\href{http://arxiv.org/abs/1705.02358}{{\ttfamily arXiv:1705.02358 [hep-ph]}}.

\bibitem{morrissey}
D.~E. Morrissey and A.~P. Spray, ``{New Limits on Light Hidden Sectors from
  Fixed-Target Experiments},''
  \href{http://dx.doi.org/10.1007/JHEP06(2014)083}{{\em JHEP} {\bfseries 06}
  (2014) 083},
\href{http://arxiv.org/abs/1402.4817}{{\ttfamily arXiv:1402.4817 [hep-ph]}}.

\bibitem{Feldman:2007wj}
D.~Feldman, Z.~Liu, and P.~Nath, ``{The Stueckelberg Z-prime Extension with
  Kinetic Mixing and Milli-Charged Dark Matter From the Hidden Sector},''
  \href{http://dx.doi.org/10.1103/PhysRevD.75.115001}{{\em Phys. Rev.}
  {\bfseries D75} (2007) 115001},
\href{http://arxiv.org/abs/hep-ph/0702123}{{\ttfamily arXiv:hep-ph/0702123
  [HEP-PH]}}.

\bibitem{Choi:2016tkj}
S.-M. Choi, Y.-J. Kang, and H.~M. Lee, ``{On thermal production of
  self-interacting dark matter},''
  \href{http://dx.doi.org/10.1007/JHEP12(2016)099}{{\em JHEP} {\bfseries 12}
  (2016) 099},
\href{http://arxiv.org/abs/1610.04748}{{\ttfamily arXiv:1610.04748 [hep-ph]}}.

\bibitem{Batell:2009di}
B.~Batell, M.~Pospelov, and A.~Ritz, ``{Exploring Portals to a Hidden Sector
  Through Fixed Targets},''
  \href{http://dx.doi.org/10.1103/PhysRevD.80.095024}{{\em Phys. Rev.}
  {\bfseries D80} (2009) 095024},
\href{http://arxiv.org/abs/0906.5614}{{\ttfamily arXiv:0906.5614 [hep-ph]}}.

\bibitem{Bjorken:2009mm}
J.~D. Bjorken, R.~Essig, P.~Schuster, and N.~Toro, ``{New Fixed-Target
  Experiments to Search for Dark Gauge Forces},''
  \href{http://dx.doi.org/10.1103/PhysRevD.80.075018}{{\em Phys. Rev.}
  {\bfseries D80} (2009) 075018},
\href{http://arxiv.org/abs/0906.0580}{{\ttfamily arXiv:0906.0580 [hep-ph]}}.

\bibitem{Schuster:2009au}
P.~Schuster, N.~Toro, and I.~Yavin, ``{Terrestrial and Solar Limits on
  Long-Lived Particles in a Dark Sector},''
  \href{http://dx.doi.org/10.1103/PhysRevD.81.016002}{{\em Phys. Rev.}
  {\bfseries D81} (2010) 016002},
\href{http://arxiv.org/abs/0910.1602}{{\ttfamily arXiv:0910.1602 [hep-ph]}}.

\bibitem{Essig:2009nc}
R.~Essig, P.~Schuster, and N.~Toro, ``{Probing Dark Forces and Light Hidden
  Sectors at Low-Energy e+e- Colliders},''
  \href{http://dx.doi.org/10.1103/PhysRevD.80.015003}{{\em Phys. Rev.}
  {\bfseries D80} (2009) 015003},
\href{http://arxiv.org/abs/0903.3941}{{\ttfamily arXiv:0903.3941 [hep-ph]}}.

\bibitem{Batell:2014mga}
B.~Batell, R.~Essig, and Z.~Surujon, ``{Strong Constraints on Sub-GeV Dark
  Sectors from SLAC Beam Dump E137},''
  \href{http://dx.doi.org/10.1103/PhysRevLett.113.171802}{{\em Phys. Rev.
  Lett.} {\bfseries 113} no.~17, (2014) 171802},
\href{http://arxiv.org/abs/1406.2698}{{\ttfamily arXiv:1406.2698 [hep-ph]}}.

\bibitem{Izaguirre:2015yja}
E.~Izaguirre, G.~Krnjaic, P.~Schuster, and N.~Toro, ``{Analyzing the Discovery
  Potential for Light Dark Matter},''
  \href{http://dx.doi.org/10.1103/PhysRevLett.115.251301}{{\em Phys. Rev.
  Lett.} {\bfseries 115} no.~25, (2015) 251301},
\href{http://arxiv.org/abs/1505.00011}{{\ttfamily arXiv:1505.00011 [hep-ph]}}.

\bibitem{deNiverville:2016rqh}
P.~deNiverville, C.-Y. Chen, M.~Pospelov, and A.~Ritz, ``{Light dark matter in
  neutrino beams: production modelling and scattering signatures at MiniBooNE,
  T2K and SHiP},'' \href{http://dx.doi.org/10.1103/PhysRevD.95.035006}{{\em
  Phys. Rev.} {\bfseries D95} no.~3, (2017) 035006},
\href{http://arxiv.org/abs/1609.01770}{{\ttfamily arXiv:1609.01770 [hep-ph]}}.

\bibitem{Battaglieri:2016ggd}
{\bfseries BDX} Collaboration, M.~Battaglieri {\em et~al.}, ``{Dark Matter
  Search in a Beam-Dump eXperiment (BDX) at Jefferson Lab},''
\href{http://arxiv.org/abs/1607.01390}{{\ttfamily arXiv:1607.01390 [hep-ex]}}.

\bibitem{Slatyer:2015jla}
T.~R. Slatyer, ``{Indirect dark matter signatures in the cosmic dark ages. I.
  Generalizing the bound on s-wave dark matter annihilation from Planck
  results},'' \href{http://dx.doi.org/10.1103/PhysRevD.93.023527}{{\em Phys.
  Rev.} {\bfseries D93} no.~2, (2016) 023527},
\href{http://arxiv.org/abs/1506.03811}{{\ttfamily arXiv:1506.03811 [hep-ph]}}.

\bibitem{Liu:2016cnk}
H.~Liu, T.~R. Slatyer, and J.~Zavala, ``{Contributions to cosmic reionization
  from dark matter annihilation and decay},''
  \href{http://dx.doi.org/10.1103/PhysRevD.94.063507}{{\em Phys. Rev.}
  {\bfseries D94} no.~6, (2016) 063507},
\href{http://arxiv.org/abs/1604.02457}{{\ttfamily arXiv:1604.02457
  [astro-ph.CO]}}.

\bibitem{Fradette:2017sdd}
A.~Fradette and M.~Pospelov, ``{BBN for the LHC: constraints on lifetimes of
  the Higgs portal scalars},''
\href{http://arxiv.org/abs/1706.01920}{{\ttfamily arXiv:1706.01920 [hep-ph]}}.

\bibitem{Kawasaki:2017bqm}
M.~Kawasaki, K.~Kohri, T.~Moroi, and Y.~Takaesu, ``{Revisiting Big-Bang
  Nucleosynthesis Constraints on Long-Lived Decaying Particles},''
\href{http://arxiv.org/abs/1709.01211}{{\ttfamily arXiv:1709.01211 [hep-ph]}}.

\bibitem{Holdom:1985ag}
B.~Holdom, ``{Two U(1)'s and Epsilon Charge Shifts},''
\href{http://dx.doi.org/10.1016/0370-2693(86)91377-8}{{\em Phys. Lett.}
  {\bfseries 166B} (1986) 196--198}.

\bibitem{Berger:2016vxi}
J.~Berger, K.~Jedamzik, and D.~G.~E. Walker, ``{Cosmological Constraints on
  Decoupled Dark Photons and Dark Higgs},''
  \href{http://dx.doi.org/10.1088/1475-7516/2016/11/032}{{\em JCAP} {\bfseries
  1611} (2016) 032},
\href{http://arxiv.org/abs/1605.07195}{{\ttfamily arXiv:1605.07195 [hep-ph]}}.

\bibitem{Batell:2009yf}
B.~Batell, M.~Pospelov, and A.~Ritz, ``{Probing a Secluded U(1) at
  B-factories},'' \href{http://dx.doi.org/10.1103/PhysRevD.79.115008}{{\em
  Phys. Rev.} {\bfseries D79} (2009) 115008},
\href{http://arxiv.org/abs/0903.0363}{{\ttfamily arXiv:0903.0363 [hep-ph]}}.

\bibitem{Batley:2015lha}
{\bfseries NA48/2} Collaboration, J.~R. Batley {\em et~al.}, ``{Search for the
  dark photon in $\pi^0$ decays},''
  \href{http://dx.doi.org/10.1016/j.physletb.2015.04.068}{{\em Phys. Lett.}
  {\bfseries B746} (2015) 178--185},
\href{http://arxiv.org/abs/1504.00607}{{\ttfamily arXiv:1504.00607 [hep-ex]}}.

\bibitem{Lees:2014xha}
{\bfseries BaBar} Collaboration, J.~P. Lees {\em et~al.}, ``{Search for a Dark
  Photon in $e^+e^-$ Collisions at BaBar},''
  \href{http://dx.doi.org/10.1103/PhysRevLett.113.201801}{{\em Phys. Rev.
  Lett.} {\bfseries 113} no.~20, (2014) 201801},
\href{http://arxiv.org/abs/1406.2980}{{\ttfamily arXiv:1406.2980 [hep-ex]}}.

\bibitem{Aaij:2017rft}
{\bfseries LHCb} Collaboration, R.~Aaij {\em et~al.}, ``{Search for dark
  photons produced in 13 TeV $pp$ collisions},''
\href{http://arxiv.org/abs/1710.02867}{{\ttfamily arXiv:1710.02867 [hep-ex]}}.

\bibitem{Andreas:2013lya}
S.~Andreas {\em et~al.}, ``{Proposal for an Experiment to Search for Light Dark
  Matter at the SPS},''
\href{http://arxiv.org/abs/1312.3309}{{\ttfamily arXiv:1312.3309 [hep-ex]}}.

\bibitem{Izaguirre:2014bca}
E.~Izaguirre, G.~Krnjaic, P.~Schuster, and N.~Toro, ``{Testing GeV-Scale Dark
  Matter with Fixed-Target Missing Momentum Experiments},''
  \href{http://dx.doi.org/10.1103/PhysRevD.91.094026}{{\em Phys. Rev.}
  {\bfseries D91} no.~9, (2015) 094026},
\href{http://arxiv.org/abs/1411.1404}{{\ttfamily arXiv:1411.1404 [hep-ph]}}.

\bibitem{Lees:2017lec}
{\bfseries BaBar} Collaboration, J.~P. Lees {\em et~al.}, ``{Search for
  Invisible Decays of a Dark Photon Produced in ${e}^{+}{e}^{-}$ Collisions at
  BaBar},'' \href{http://dx.doi.org/10.1103/PhysRevLett.119.131804}{{\em Phys.
  Rev. Lett.} {\bfseries 119} no.~13, (2017) 131804},
\href{http://arxiv.org/abs/1702.03327}{{\ttfamily arXiv:1702.03327 [hep-ex]}}.

\bibitem{Banerjee:2017hhz}
{\bfseries NA64} Collaboration, D.~Banerjee {\em et~al.}, ``{Search for vector
  mediator of Dark Matter production in invisible decay mode},''
\href{http://arxiv.org/abs/1710.00971}{{\ttfamily arXiv:1710.00971 [hep-ex]}}.

\bibitem{Feroz:2008xx}
F.~Feroz, M.~P. Hobson, and M.~Bridges, ``{MultiNest: an efficient and robust
  Bayesian inference tool for cosmology and particle physics},''
  \href{http://dx.doi.org/10.1111/j.1365-2966.2009.14548.x}{{\em Mon. Not. Roy.
  Astron. Soc.} {\bfseries 398} (2009) 1601--1614},
\href{http://arxiv.org/abs/0809.3437}{{\ttfamily arXiv:0809.3437 [astro-ph]}}.

\bibitem{Ade:2015xua}
{\bfseries Planck} Collaboration, P.~A.~R. Ade {\em et~al.}, ``{Planck 2015
  results. XIII. Cosmological parameters},''
  \href{http://dx.doi.org/10.1051/0004-6361/201525830}{{\em Astron. Astrophys.}
  {\bfseries 594} (2016) A13},
\href{http://arxiv.org/abs/1502.01589}{{\ttfamily arXiv:1502.01589
  [astro-ph.CO]}}.

\bibitem{Belanger:2014vza}
G.~Bélanger, F.~Boudjema, A.~Pukhov, and A.~Semenov, ``{micrOMEGAs4.1: two
  dark matter candidates},''
  \href{http://dx.doi.org/10.1016/j.cpc.2015.03.003}{{\em Comput. Phys.
  Commun.} {\bfseries 192} (2015) 322--329},
\href{http://arxiv.org/abs/1407.6129}{{\ttfamily arXiv:1407.6129 [hep-ph]}}.

\bibitem{Porod:2003um}
W.~Porod, ``{SPheno, a program for calculating supersymmetric spectra, SUSY
  particle decays and SUSY particle production at e+ e- colliders},''
  \href{http://dx.doi.org/10.1016/S0010-4655(03)00222-4}{{\em Comput. Phys.
  Commun.} {\bfseries 153} (2003) 275--315},
\href{http://arxiv.org/abs/hep-ph/0301101}{{\ttfamily arXiv:hep-ph/0301101
  [hep-ph]}}.

\bibitem{Porod:2011nf}
W.~Porod and F.~Staub, ``{SPheno 3.1: Extensions including flavour, CP-phases
  and models beyond the MSSM},''
  \href{http://dx.doi.org/10.1016/j.cpc.2012.05.021}{{\em Comput. Phys.
  Commun.} {\bfseries 183} (2012) 2458--2469},
\href{http://arxiv.org/abs/1104.1573}{{\ttfamily arXiv:1104.1573 [hep-ph]}}.

\bibitem{staub_sarah_2008}
F.~Staub, ``{SARAH},''
\href{http://arxiv.org/abs/0806.0538}{{\ttfamily arXiv:0806.0538 [hep-ph]}}.

\bibitem{Staub:2012pb}
F.~Staub, ``{SARAH 3.2: Dirac Gauginos, UFO output, and more},''
  \href{http://dx.doi.org/10.1016/j.cpc.2013.02.019}{{\em Comput. Phys.
  Commun.} {\bfseries 184} (2013) 1792--1809},
\href{http://arxiv.org/abs/1207.0906}{{\ttfamily arXiv:1207.0906 [hep-ph]}}.

\bibitem{Staub:2013tta}
F.~Staub, ``{SARAH 4 : A tool for (not only SUSY) model builders},''
  \href{http://dx.doi.org/10.1016/j.cpc.2014.02.018}{{\em Comput. Phys.
  Commun.} {\bfseries 185} (2014) 1773--1790},
\href{http://arxiv.org/abs/1309.7223}{{\ttfamily arXiv:1309.7223 [hep-ph]}}.

\bibitem{Liu:2014cma}
J.~Liu, N.~Weiner, and W.~Xue, ``{Signals of a Light Dark Force in the Galactic
  Center},'' \href{http://dx.doi.org/10.1007/JHEP08(2015)050}{{\em JHEP}
  {\bfseries 08} (2015) 050},
\href{http://arxiv.org/abs/1412.1485}{{\ttfamily arXiv:1412.1485 [hep-ph]}}.

\bibitem{Gondolo:1990dk}
P.~Gondolo and G.~Gelmini, ``{Cosmic abundances of stable particles: Improved
  analysis},''
\href{http://dx.doi.org/10.1016/0550-3213(91)90438-4}{{\em Nucl. Phys.}
  {\bfseries B360} (1991) 145--179}.

\bibitem{Kolb:1990vq}
E.~W. Kolb and M.~S. Turner, ``{The Early Universe},''
{\em Front. Phys.} {\bfseries 69} (1990) 1--547.

\bibitem{Boehm:2003hm}
C.~Boehm and P.~Fayet, ``{Scalar dark matter candidates},''
  \href{http://dx.doi.org/10.1016/j.nuclphysb.2004.01.015}{{\em Nucl. Phys.}
  {\bfseries B683} (2004) 219--263},
\href{http://arxiv.org/abs/hep-ph/0305261}{{\ttfamily arXiv:hep-ph/0305261
  [hep-ph]}}.

\bibitem{deNiverville:2011it}
P.~deNiverville, M.~Pospelov, and A.~Ritz, ``{Observing a light dark matter
  beam with neutrino experiments},''
  \href{http://dx.doi.org/10.1103/PhysRevD.84.075020}{{\em Phys. Rev.}
  {\bfseries D84} (2011) 075020},
\href{http://arxiv.org/abs/1107.4580}{{\ttfamily arXiv:1107.4580 [hep-ph]}}.

\bibitem{Pospelov:2007mp}
M.~Pospelov, A.~Ritz, and M.~B. Voloshin, ``{Secluded WIMP Dark Matter},''
  \href{http://dx.doi.org/10.1016/j.physletb.2008.02.052}{{\em Phys. Lett.}
  {\bfseries B662} (2008) 53--61},
\href{http://arxiv.org/abs/0711.4866}{{\ttfamily arXiv:0711.4866 [hep-ph]}}.

\bibitem{Kouvaris:2014uoa}
C.~Kouvaris, I.~M. Shoemaker, and K.~Tuominen, ``{Self-Interacting Dark Matter
  through the Higgs Portal},''
  \href{http://dx.doi.org/10.1103/PhysRevD.91.043519}{{\em Phys. Rev.}
  {\bfseries D91} no.~4, (2015) 043519},
\href{http://arxiv.org/abs/1411.3730}{{\ttfamily arXiv:1411.3730 [hep-ph]}}.

\bibitem{Krnjaic:2015mbs}
G.~Krnjaic, ``{Probing Light Thermal Dark-Matter With a Higgs Portal
  Mediator},'' \href{http://dx.doi.org/10.1103/PhysRevD.94.073009}{{\em Phys.
  Rev.} {\bfseries D94} no.~7, (2016) 073009},
\href{http://arxiv.org/abs/1512.04119}{{\ttfamily arXiv:1512.04119 [hep-ph]}}.

\bibitem{Izaguirre:2017bqb}
E.~Izaguirre, Y.~Kahn, G.~Krnjaic, and M.~Moschella, ``{Testing Light Dark
  Matter Coannihilation With Fixed-Target Experiments},''
  \href{http://dx.doi.org/10.1103/PhysRevD.96.055007}{{\em Phys. Rev.}
  {\bfseries D96} no.~5, (2017) 055007},
\href{http://arxiv.org/abs/1703.06881}{{\ttfamily arXiv:1703.06881 [hep-ph]}}.

\bibitem{Nollett:2013pwa}
K.~M. Nollett and G.~Steigman, ``{BBN And The CMB Constrain Light,
  Electromagnetically Coupled WIMPs},''
  \href{http://dx.doi.org/10.1103/PhysRevD.89.083508}{{\em Phys. Rev.}
  {\bfseries D89} no.~8, (2014) 083508},
\href{http://arxiv.org/abs/1312.5725}{{\ttfamily arXiv:1312.5725
  [astro-ph.CO]}}.

\bibitem{Tiffenberg:2017aac}
J.~Tiffenberg, M.~Sofo-Haro, A.~Drlica-Wagner, R.~Essig, Y.~Guardincerri,
  S.~Holland, T.~Volansky, and T.-T. Yu, ``{Single-electron and single-photon
  sensitivity with a silicon Skipper CCD},''
  \href{http://dx.doi.org/10.1103/PhysRevLett.119.131802}{{\em Phys. Rev.
  Lett.} {\bfseries 119} no.~13, (2017) 131802},
\href{http://arxiv.org/abs/1706.00028}{{\ttfamily arXiv:1706.00028
  [physics.ins-det]}}.

\bibitem{Essig:2017kqs}
R.~Essig, T.~Volansky, and T.-T. Yu, ``{New Constraints and Prospects for
  sub-GeV Dark Matter Scattering off Electrons in Xenon},''
  \href{http://dx.doi.org/10.1103/PhysRevD.96.043017}{{\em Phys. Rev.}
  {\bfseries D96} no.~4, (2017) 043017},
\href{http://arxiv.org/abs/1703.00910}{{\ttfamily arXiv:1703.00910 [hep-ph]}}.

\bibitem{Essig:2012yx}
R.~Essig, A.~Manalaysay, J.~Mardon, P.~Sorensen, and T.~Volansky, ``{First
  Direct Detection Limits on sub-GeV Dark Matter from XENON10},''
  \href{http://dx.doi.org/10.1103/PhysRevLett.109.021301}{{\em Phys. Rev.
  Lett.} {\bfseries 109} (2012) 021301},
\href{http://arxiv.org/abs/1206.2644}{{\ttfamily arXiv:1206.2644
  [astro-ph.CO]}}.

\bibitem{Kaplinghat:2013yxa}
M.~Kaplinghat, S.~Tulin, and H.-B. Yu, ``{Direct Detection Portals for
  Self-interacting Dark Matter},''
  \href{http://dx.doi.org/10.1103/PhysRevD.89.035009}{{\em Phys. Rev.}
  {\bfseries D89} no.~3, (2014) 035009},
\href{http://arxiv.org/abs/1310.7945}{{\ttfamily arXiv:1310.7945 [hep-ph]}}.

\bibitem{Athanassopoulos:1996ds}
{\bfseries LSND} Collaboration, C.~Athanassopoulos {\em et~al.}, ``{The Liquid
  scintillator neutrino detector and LAMPF neutrino source},''
  \href{http://dx.doi.org/10.1016/S0168-9002(96)01155-2}{{\em Nucl. Instrum.
  Meth.} {\bfseries A388} (1997) 149--172},
\href{http://arxiv.org/abs/nucl-ex/9605002}{{\ttfamily arXiv:nucl-ex/9605002
  [nucl-ex]}}.

\bibitem{AguilarArevalo:2008qa}
{\bfseries MiniBooNE} Collaboration, A.~A. Aguilar-Arevalo {\em et~al.}, ``{The
  MiniBooNE Detector},''
  \href{http://dx.doi.org/10.1016/j.nima.2008.10.028}{{\em Nucl. Instrum.
  Meth.} {\bfseries A599} (2009) 28--46},
\href{http://arxiv.org/abs/0806.4201}{{\ttfamily arXiv:0806.4201 [hep-ex]}}.

\bibitem{Antonello:2015lea}
{\bfseries LAr1-ND, ICARUS-WA104, MicroBooNE} Collaboration, M.~Antonello {\em
  et~al.}, ``{A Proposal for a Three Detector Short-Baseline Neutrino
  Oscillation Program in the Fermilab Booster Neutrino Beam},''
\href{http://arxiv.org/abs/1503.01520}{{\ttfamily arXiv:1503.01520
  [physics.ins-det]}}.

\bibitem{Riordan:1987aw}
E.~M. Riordan {\em et~al.}, ``{A Search for Short Lived Axions in an Electron
  Beam Dump Experiment},''
\href{http://dx.doi.org/10.1103/PhysRevLett.59.755}{{\em Phys. Rev. Lett.}
  {\bfseries 59} (1987) 755}.

\bibitem{Batell:2014yra}
B.~Batell, P.~deNiverville, D.~McKeen, M.~Pospelov, and A.~Ritz, ``{Leptophobic
  Dark Matter at Neutrino Factories},''
  \href{http://dx.doi.org/10.1103/PhysRevD.90.115014}{{\em Phys. Rev.}
  {\bfseries D90} no.~11, (2014) 115014},
\href{http://arxiv.org/abs/1405.7049}{{\ttfamily arXiv:1405.7049 [hep-ph]}}.

\bibitem{Athanassopoulos:1997er}
{\bfseries LSND} Collaboration, C.~Athanassopoulos {\em et~al.}, ``{Evidence
  for muon-neutrino $\rightarrow$ electron-neutrino oscillations from pion
  decay in flight neutrinos},''
  \href{http://dx.doi.org/10.1103/PhysRevC.58.2489}{{\em Phys. Rev.} {\bfseries
  C58} (1998) 2489--2511},
\href{http://arxiv.org/abs/nucl-ex/9706006}{{\ttfamily arXiv:nucl-ex/9706006
  [nucl-ex]}}.

\bibitem{Aguilar:2001ty}
{\bfseries LSND} Collaboration, A.~Aguilar-Arevalo {\em et~al.}, ``{Evidence
  for neutrino oscillations from the observation of anti-neutrino(electron)
  appearance in a anti-neutrino(muon) beam},''
  \href{http://dx.doi.org/10.1103/PhysRevD.64.112007}{{\em Phys. Rev.}
  {\bfseries D64} (2001) 112007},
\href{http://arxiv.org/abs/hep-ex/0104049}{{\ttfamily arXiv:hep-ex/0104049
  [hep-ex]}}.

\bibitem{Aguilar-Arevalo:2017mqx}
{\bfseries MiniBooNE} Collaboration, A.~A. Aguilar-Arevalo {\em et~al.},
  ``{Dark Matter Search in a Proton Beam Dump with MiniBooNE},''
  \href{http://dx.doi.org/10.1103/PhysRevLett.118.221803}{{\em Phys. Rev.
  Lett.} {\bfseries 118} no.~22, (2017) 221803},
\href{http://arxiv.org/abs/1702.02688}{{\ttfamily arXiv:1702.02688 [hep-ex]}}.

\bibitem{Patterson:2009ki}
R.~B. Patterson, E.~M. Laird, Y.~Liu, P.~D. Meyers, I.~Stancu, and H.~A.
  Tanaka, ``{The Extended-track reconstruction for MiniBooNE},''
  \href{http://dx.doi.org/10.1016/j.nima.2009.06.064}{{\em Nucl. Instrum.
  Meth.} {\bfseries A608} (2009) 206--224},
\href{http://arxiv.org/abs/0902.2222}{{\ttfamily arXiv:0902.2222 [hep-ex]}}.

\bibitem{Cyburt:2002uv}
R.~H. Cyburt, J.~R. Ellis, B.~D. Fields, and K.~A. Olive, ``{Updated
  nucleosynthesis constraints on unstable relic particles},''
  \href{http://dx.doi.org/10.1103/PhysRevD.67.103521}{{\em Phys. Rev.}
  {\bfseries D67} (2003) 103521},
\href{http://arxiv.org/abs/astro-ph/0211258}{{\ttfamily arXiv:astro-ph/0211258
  [astro-ph]}}.

\bibitem{Kawasaki:2004yh}
M.~Kawasaki, K.~Kohri, and T.~Moroi, ``{Hadronic decay of late - decaying
  particles and Big-Bang Nucleosynthesis},''
  \href{http://dx.doi.org/10.1016/j.physletb.2005.08.045}{{\em Phys. Lett.}
  {\bfseries B625} (2005) 7--12},
\href{http://arxiv.org/abs/astro-ph/0402490}{{\ttfamily arXiv:astro-ph/0402490
  [astro-ph]}}.

\bibitem{Jedamzik:2006xz}
K.~Jedamzik, ``{Big bang nucleosynthesis constraints on hadronically and
  electromagnetically decaying relic neutral particles},''
  \href{http://dx.doi.org/10.1103/PhysRevD.74.103509}{{\em Phys. Rev.}
  {\bfseries D74} (2006) 103509},
\href{http://arxiv.org/abs/hep-ph/0604251}{{\ttfamily arXiv:hep-ph/0604251
  [hep-ph]}}.

\bibitem{Pospelov:2010hj}
M.~Pospelov and J.~Pradler, ``{Big Bang Nucleosynthesis as a Probe of New
  Physics},'' \href{http://dx.doi.org/10.1146/annurev.nucl.012809.104521}{{\em
  Ann. Rev. Nucl. Part. Sci.} {\bfseries 60} (2010) 539--568},
\href{http://arxiv.org/abs/1011.1054}{{\ttfamily arXiv:1011.1054 [hep-ph]}}.

\bibitem{planck2015}
{\bfseries Planck} Collaboration, P.~A.~R. Ade {\em et~al.}, ``{Planck 2015
  results. XIII. Cosmological parameters},''
  \href{http://dx.doi.org/10.1051/0004-6361/201525830}{{\em Astron. Astrophys.}
  {\bfseries 594} (2016) A13},
\href{http://arxiv.org/abs/1502.01589}{{\ttfamily arXiv:1502.01589
  [astro-ph.CO]}}.

\bibitem{Smolinsky:2017fvb}
J.~Smolinsky and P.~Tanedo, ``{Dark Photons from Captured Inelastic Dark Matter
  Annihilation: Charged Particle Signatures},''
  \href{http://dx.doi.org/10.1103/PhysRevD.95.075015}{{\em Phys. Rev.}
  {\bfseries D95} no.~7, (2017) 075015},
\href{http://arxiv.org/abs/1701.03168}{{\ttfamily arXiv:1701.03168 [hep-ph]}}.

\bibitem{Kahn:2014sra}
Y.~Kahn, G.~Krnjaic, J.~Thaler, and M.~Toups, ``{DAEδALUS and dark matter
  detection},'' \href{http://dx.doi.org/10.1103/PhysRevD.91.055006}{{\em Phys.
  Rev.} {\bfseries D91} no.~5, (2015) 055006},
\href{http://arxiv.org/abs/1411.1055}{{\ttfamily arXiv:1411.1055 [hep-ph]}}.

\end{thebibliography}\endgroup


\providecommand{\href}[2]{#2}\begingroup\raggedright\endgroup

\end{document}